\documentclass[times,twocolumn,final]{elsarticle}

\usepackage{jasr}
\usepackage{framed,multirow}
\usepackage{amssymb}
\usepackage{latexsym}
\usepackage[switch]{lineno}
\usepackage{url}
\usepackage{xcolor}
\definecolor{newcolor}{rgb}{.8,.349,.1}
\usepackage{algpseudocode}
\usepackage{algorithm}

\usepackage{hyperref}
\hypersetup{
  colorlinks   = true, 
  urlcolor     = cyan, 
  linkcolor    = cyan, 
  citecolor   = red 
}
\newcommand{\comment}[1]{}

\newcommand{\scitechfig}{Reproduced from \citet{bhaskara2025development}.}
\usepackage[normalem]{ulem}
\usepackage{mathtools}
\usepackage{amsmath}
\usepackage{subfig}
\usepackage{tabularx}
\usepackage[title]{appendix}
\usepackage{placeins}
\usepackage{adjustbox}
\usepackage{enumitem}
\usepackage{scrextend} 

\makeatletter
\def\stmtitleheader{}
\renewenvironment{abstract}
  {\global\setbox\absbox=\vtop\bgroup\hsize=\textwidth\bgroup\elsabsfont\unskip\ignorespaces}
  {\egroup\egroup}
\def\ps@headings{%
  \def\@evenhead{{\titleheaderfont\thepage\hfill}}%
  \def\@oddhead{{\titleheaderfont\hfill\thepage}}%
  \let\@oddfoot\@empty
  \let\@evenfoot\@empty}
\makeatother

\begin{document}

\verso{R.R Bhaskara \textit{et al.}}

\begin{frontmatter}

\title{NaRPA: Navigation and Rendering Pipeline for Astronautics}

\author[1]{Ramchander Rao {Bhaskara}\corref{cor1}}
\cortext[cor1]{Equal contribution.}
\ead{bhaskara@tamu.edu}
\author[2]{Roshan Thomas Eapen\corref{cor1}}
\ead{rpe5185@psu.edu}
\author[1]{Manoranjan Majji}
\ead{mmajji@tamu.edu}

\affiliation[1]{organization={Department of Aerospace Engineering, Texas A\&M University},
                addressline={710 Ross St},
                city={College Station},
                postcode={TX 77843},
                country={USA}}

\affiliation[2]{organization={Department of Aerospace Engineering, Pennsylvania State University},
                addressline={229 Hammond Building},
                city={University Park},
                postcode={PA 16802},
                country={USA}}


\begin{abstract}
This paper presents the applications of scientific ray-tracing in modeling and simulating light transport for space-borne image data generation. 
A ray-tracing engine, the Navigation and Rendering Pipeline for Astronautics (NaRPA), is introduced as a rendering framework to generate virtual datasets and support simulations for robust navigation pipelines.
Sensor and environment models that enable the synthesis of space-to-space and ground-to-space virtual observations are presented.  
The work demonstrates the capabilities of simulating passive and active vision-based sensors using NaRPA to facilitate the design, testing, and verification of aerospace visual navigation algorithms. 
Additionally, the paper describes a velocimeter LiDAR model and its statistical validation with experimental data.
\end{abstract}

\end{frontmatter}


\section{Introduction} \label{intro}

High-fidelity modeling and simulations are critical in the successful realization of complex spaceflight endeavors, encompassing both robotic and manned missions. 
At the core of these advanced simulations lies the intricate modeling of imaging physics, which has become instrumental in fostering novel approaches to vision-based navigation in spacecraft guidance. 
Optical sensing and image processing techniques significantly influence a wide spectrum of aerospace applications, including terrain relative navigation and hazard avoidance~\citep{mccabe2020anonymous,johnson2023implementation}, on-orbit servicing~\citep{zhang2008vision}, rendezvous and docking operations~\citep{liu2014relative}, and debris removal missions~\citep{forshaw2016removedebris}. 
The development and verification of such optical sensing applications heavily rely on fidelity of visual simulations~\citep{penn2016trick}, underscoring the critical need for rendering physically realistic data to synthesize empirically-valid space-based observations.

Given the intricate relationship between image synthesis and visual navigation, there exists a need for rendering tools that accurately model illumination in free space. 
This requirement is crucial for developing robust navigation algorithms that can perform reliably in the challenging conditions of space missions.
This paper explores key aspects of image formation essential for modeling physically accurate space scene simulations. 
The Navigation and Rendering Pipeline for Astronautics (NaRPA), an image rendering engine developed based on standard computer graphics principles, is introduced and is shown to support optical sensing for accurate visual data synthesis in aerospace mission applications. 
Through NaRPA, key aspects of visual data generation pipelines and model applications are demonstrated, indicating the utility of high-fidelity rendering in planetary remote sensing.

Image rendering methods are developed based on two complementary approaches: {ray tracing and rasterization.
A ray tracing algorithm loops over all the pixels in the image; casts a ray into the scene for each pixel to compute their coloration~\citep{glassner1989introduction}.
A rasterization algorithm instead loops over all the objects in the scene, computing for each object the pixels covered by that object~\citep{hughes2014computer}. The resulting per-object pixels are formatted for a raster display. 
Rasterization, a core component in graphical programming interfaces such as OpenGL~\citep{shreiner2013opengl} involves projection of 3D objects in a scene to generate a two-dimensional image~\citep{pineda1988parallel}. 
While this technique has a low computational cost, handling of global effects of light such as reflections and shading, is primitive~\citep{davidovivc20123d}. These global effects make ray tracing a competitive choice for rendering {photorealistic} images. 


The basic concepts of ray tracing were developed and popularized by~\cite{whitted1979improved}. Expanded into probabilistic ray-tracing,~\cite{kajiya1986rendering} proposed the {rendering equation} that provides a mathematical description of the light energy distribution in a scene. The rendering equation is computationally very expensive because, for every point on the surface of an object, it also evaluates the effects of geometry and scattering properties of all other objects in the scene. Rendering algorithms that capture this complexity are called {global illumination} algorithms, which generally implement finite element or Monte Carlo methods to solve the rendering equation. 
Monte Carlo techniques account for complex scene geometries and light-matter interactions allowing for accurate simulation of phenomena like multiple scattering, specular reflections, and shadows from irregularly shaped objects such as spacecraft or asteroids.
Key techniques include bidirectional path tracing~\citep{10.1007/978-3-7091-7484-5_10} which improve convergence and reduce noise in the rendered image. The main challenge in space environment modeling is balancing accuracy with computational efficiency, as simulating the vast scales and complex light behaviors in space often requires significant computational resources. Recent advancements in hardware acceleration and optimized algorithms~\citep{parker2013gpu} have made real-time or near-real-time ray tracing more feasible for space environment simulations.

Software platforms to realize the ray tracing architecture have adapted to embed the incremental development in research. These efforts are directed at rendering photorealistic images with optimal computational costs. 
Radiance is one of the first open source rendering software that fundamentally uses ray-tracing techniques to compute pixel illumination values to render an image that provides scientifically validated lighting simulations~\citep{ward1994radiance}.
Early rendering systems such as Spectrum~\citep{glassner1993spectrum} and Vision~\citep{slusallek1995vision} provided incremental development to robust software realizations of physically based rendering engines through a formulation of a structure for object-scene description. Development of other ray tracing algorithms (PBRT~\citep{10.5555/3044800}, POV Ray~\citep{plachetka1998pov}, Blender Cycles~\citep{10.5555/2800545}, OptiX~\citep{parker2010optix}) aided rendering of complex geometries and environments become more and more practical at modest computational cost on most hardware platforms. Rendering system efforts such as {Mitsuba~\citep{jakob2013mitsuba} and Digital Imaging and Remote Sensing Image Generation (DIRSIG)~\citep{goodenough2012dirsig} are targeted for research applications to model physically accurate light transport phenomenon. The proposed contribution is inspired from the strong emphasis on experimental rendering attempted by Mitsuba.} {Even though the state-of-the-art rendering engines could elegantly capture photorealistic effects, pipelines to model multi-sensor space-borne navigation capabilities under varying lighting conditions are desired. 
This research is aimed at applications of a simulation engine that realistically embeds pipelines for various space-borne navigation scenarios with an accessible approach.

A tremendous amount engineering effort has put into development of high-fidelity simulation software for planetary robotic exploration. 
Much of these independent efforts have been developed with focus on requirements of respective space agencies and are usually license constrained.    
For instance, PANGU~\citep{parkes2004planet}, SurRender~\citep{brochard2018scientific}, and DARTS~\citep{aiazzi2022iris} are specialized efforts on terrain and sensor modeling for astronautic simulations respectively developed for mission designs at ESA, Airbus, and NASA.
Alongside radiometrically calibrated camera and LiDAR models, high-fidelity software implement SPICE\footnote{SPICE ancillary information toolkit: \url{https://naif.jpl.nasa.gov/naif/index.html}, accessed on August 5, 2024.\label{fn:SPICE}} (Spacecraft, Planet, Instrument, C-matrix, Events) toolkit~\citep{martin2021pangu} which provides observation geometric metadata of the pointing instrument and the position of the spacecraft needed to simulate realistic remote sensing trajectories. 
PANGU also implements a single scattering atmospheric model for simulations involving Earth, or Mars., extending simulations beyond free-space. 
SISPO~\citep{pajusalu2022sispo} is developed to provide a pipeline for simulated imagery in proximity operations and is based on Blender Cycles~\citep{10.5555/2800545} for rendering. 
Photorealistic engines such as Blender and Unreal Engine are shown to propound the attributes of a generalized graphics tool as a space scene generation utility.
These tools enable development of custom shaders and procedural methods that cater to the simulation needs of specific planetary bodies under respective studies~\citep{villa2023image}.
CORTO~\citep{pugliatti2023corto} and MONET~\citep{buonagura2024monet} are developed for asteroid surface simulations, NASA's DUST~\citep{bingham2023digital} and Astrobotic's LunaRay~\citep{hough2022demkit} are created for lunar surface operations, and GUISS~\citep{bhaskara2024icy} is developed for the visualization of icy-moon terrains. 
These studies indicate the challenges and efforts in rendering physically accurate space scenes for engineering representation.

Based on these challenges the attributes of a visual sensor simulator for astronautic navigation shall consist of: (1) a full pipeline for simulation of trajectories for landing or rendezvous operations for physically accurate environment modeling, (2) implementation of user configurable sensor and lighting models in a virtual environment to render images on the fly, while being sensitive to the atmospheric scattering effects for ground-based space-object observations, (3) a powerful and accessible utility in providing ground truth data and a simulation environment for space-borne visual navigation. 
This paper illustrates these characteristics through the model graphical simulation platform, NaRPA, developed based on standard ray-tracing principles and originally envisioned for accurate entry, descent, and landing (EDL) sensor simulations. 
Embedding a heterogenous sensor package in a navigation pipeline, applications are put forth to emphasize the utility of ground truthing and testing platform in visual navigation.
Building upon the imaging capabilities of NaRPA demonstrated through our previous projects on terrain relative navigation~\citep{verras2021vision} and pose refinement~\citep{bhaskara2023differentiable}, attention in this work is given to validation of the proposed velocimeter LiDAR\footnote{Velocimeter LiDAR refers to a range measurement device that can measure the 3D position of a point it sees in space and the corresponding Doppler velocity for each such point. It also referred to as Frequency Modulated Carrier Wave (FMCW) LiDAR or Doppler LiDAR.\label{fn:velocimeterLIDAR}} model from laboratory experiments.

\comment{
the deployment of the NaRPA engine for the generation of stereoscopic images and the estimation of 3D coordinates. Another key application of NaRPA is to differentiably render images on-the-fly using feedback from vision sensors to simulate spacecraft guidance laws for proximity operations. Differentiable rendering technique enables relative pose estimation based on error in image formation. As opposed to error in point features, this error includes illumination, material properties, and other artifacts that cannot be captured using constant illumination models.

}


 \comment{
 In summary, this paper discusses ray-tracing as an effective simulation tool for vision-based navigation in aerospace missions. It presents the principles of ray tracing to highlight the development of a physically-based engine with space scene rendering capabilities.  
 Embedding a heterogenous sensor package in a navigation pipeline, applications are put forth to emphasize the utility of ground truthing and testing platform in visual navigation.  
 
 presents the development of NaRPA, a research-oriented rendering engine specialized for navigation of aerospace vehicles. NaRPA enables the modeling of ground and space-based optical images and point-clouds. It seeks to address standing gaps in space-borne image synthesis with careful attention to geometry, scaling, illumination, and scene handling. The rendering engine described in this paper emphasizes rendering lighting simulations that are physically accurate to within ray-optics model. NaRPA has the capability to simulate parameterizable camera, LiDAR, and depth sensor observations to seamlessly render images, point clouds, and depth maps, respectively. The engine includes a complete navigation pipeline and generates continuous vision-based measurements from heterogenous sensor systems that are true to their physical models and are sensitive to the environmental phenomena. Statistical methods to estimate relative pose from simulated measurements are also implemented.}
 
 This paper is organized as follows. In Section~\ref{sec:s2_pipeline}, the ray-tracing based graphics pipeline is presented. In Section~\ref{sec:s3_RenderingEq}, the mathematical setup of the rendering equation, the lighting model, and the acceleration structure for solving the rendering equation are described. Various capabilities of ray-tracing in space-borne imaging are also highlighted in this section. In Section~\ref{sec:atmosphericModel}, an atmospheric modeling procedure that simulates the effects of light scattering phenomenon is discussed, and the simulation results are presented with representative applications focused on ground-based space-object observations. In Section~\ref{sec:velocimeterLidar}, statistical validation of the velocimeter LiDAR model is presented. Limitations and future directions are discussed in Section~\ref{sec:s5_discussion}. Concluding remarks are made in Section~\ref{sec:s6_conclusion}. 	  
 
\section{Rendering Pipeline} \label{sec:s2_pipeline}

\comment{
\textit{Rendering} is the process of synthesizing images from a geometric description of a virtual environment or a \textit{scene} using a computer application. The software framework that invokes certain hardware capabilities to synthesize an image is a \textit{rendering system} or a \textit{rendering engine} or a \textit{ray tracer engine}. This framework is a collection of abstract base classes and functions that are run together to translate the said geometric description of a three-dimensional scene to the image space. A rendering system such as the one presented in this paper is a specialized and standalone application with interfaces that enable the system to render images based on a user-provided description of a virtual scene. 
The sequence of steps followed by the rendering system in order to render a scene geometry is collectively defined in its \textit{graphics} or \textit{rendering pipeline}. The stages of the graphics pipeline are implemented on hardware but are generally accessible through a graphics application programming interface (API). The rendering system presented in this paper is programmed as a portable C++ toolkit which also provides a simple high-level graphics API for scene description. In the following section, we introduce the components of the rendering pipeline that build the proposed rendering system.

\subsection{The ray-tracing graphics pipeline}
}
 \begin{figure*}[t]
        \centering
        \includegraphics[width=\textwidth]{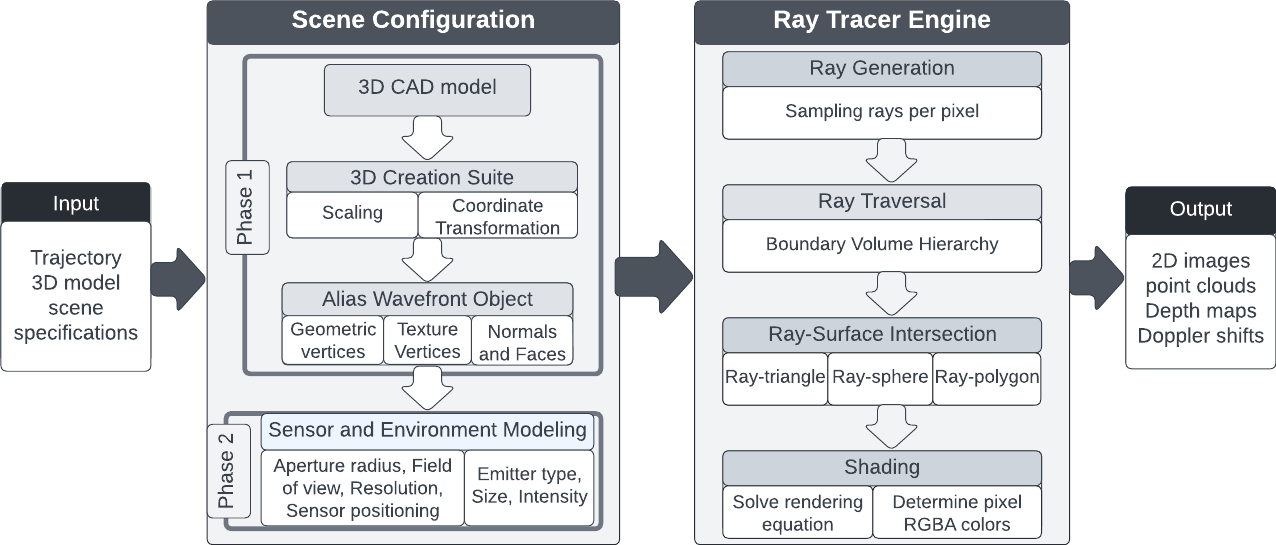}
        \caption{Architecture of Navigation and Rendering Pipeline (NaRPA).}
        \comment{\caption{Flow diagram illustrating the steps involved in the rendering pipeline. The scene configuration block of the rendering pipeline sets up the 3D environment and the sensor parameters. The ray tracer engine generates rays for the scene traversal and compute illumination at each of the ray-surface intersections to produce the final image.}}
         \label{fig:renderingPipeline}
  \end{figure*}

The NaRPA rendering engine presented in this work is a standalone application with interfaces that enable the system to render images based on a user-provided description of a virtual scene. 
NaRPA is a portable C++ toolkit which also provides a simple high-level graphics application programming interface for scene description. In this section, the components of the rendering pipeline are introduced.

The framework that constitutes the flow of the NaRPA's rendering system pipeline is illustrated in Fig.~\ref{fig:renderingPipeline}. The rendering pipeline involves two  blocks: a) {scene configuration}, and b) a {ray tracer engine}. 
The scene configuration block provides to the ray tracer, a description of the scene geometry, lighting, camera, and other attributes that are required to generate an image of user-specified configuration. The renderer invokes the ray tracing algorithm and calculates the pixel colors by solving the rendering equation. 
The rendering system is designed to handle these two specialized blocks in order to render a virtual scene. Each of the processes in Fig.~\ref{fig:renderingPipeline} is explained below.

\subsubsection{Scene configuration}
Computerized 3D geometry of an object is conventionally built from fundamental polygonal elements such as triangles, quadrilaterals, hexagons, or spheres. The 3D geometry is defined from the coordinates that correspond to geometric vertices, texture vertices (for objects with texture overlayed), vertex normals, and faces that make each polygon. Each of these vertices are indexed within an Alias Wavefront Object (.obj) file~\citep{rose1993wavefront}. The Wavefront object (OBJ) file is a standard file format adopted to define 3D geometry for surfaces containing one or more objects. In addition, an OBJ file also invokes a Material Library File (MTL) to describe colors and textures of the 3D surfaces. The scene configuration component of the rendering system accepts a 3D CAD model of the scene or the objects in a scene. 3D creation suites such as  Blender~\citep{hess2013blender2}  or MeshLab~\citep{Cignoni2008MeshLabAO} can be used to scale the 3D scene geometry as well as position it in the 3D world environment. They help convert the CAD model into the OBJ file format, which is then fed to the ray-tracer engine.  

With reference to Fig.~\ref{fig:renderingPipeline}, the scene configuration block is divided into a two phased routine to complete the scene description. Manufacturing the 3D scene in the OBJ file format completes the first part of the routine. The second phase requires a user specification of the sensor (for example, camera parameters: aperture, focus distance, field-of-view) and the emitter source parameters (light source, sun position, etc.) to respectively describe the image output requirements and the lighting environment. The ray tracer engine accepts the scene description in a predefined scene handling framework which defines the overall scene geometry, emitter sources, sensor parameters, and {additional user configurable surrounding environment such as} atmospheric properties, for the rendering system to begin the image synthesis process. The scene configuration framework bridges  the  3D  scene  infrastructure  and  the  rendering requirements.

\begin{figure}[t]
        \centering
        \includegraphics[width=0.48\textwidth]{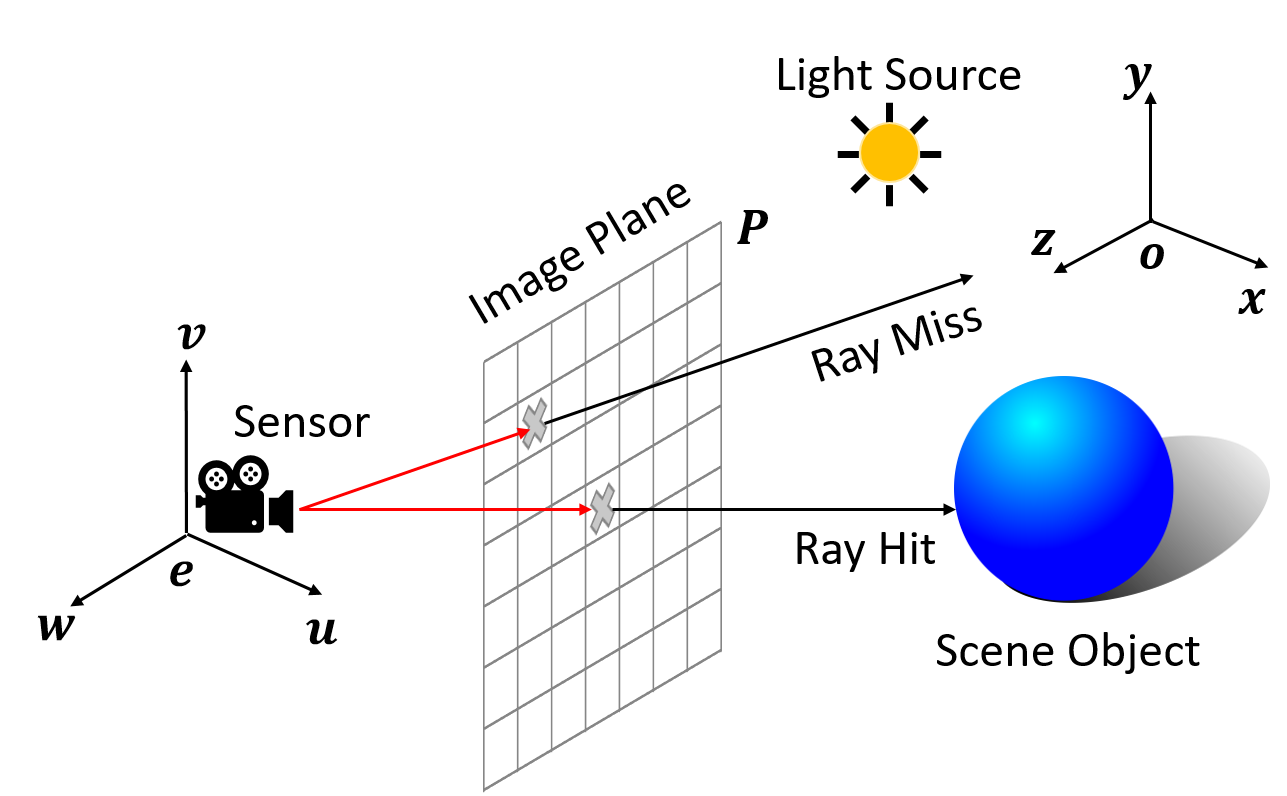}
        \caption{Ray-tracing geometry.}
        \comment{
        \caption{The ray tracing geometry: A ray cast from the sensor (camera), through the image plane, intersects the scene object in its course. The scene geometry and the light source (Sun) is set up in the $xyz$ frame of reference, whereas the camera is aligned to a $uvw$ coordinate frame.}
        }
         \label{fig:viewingGeometry}
        \end{figure}

\subsubsection{Ray tracing}
The ray tracing engine manages the rendering process by executing the rendering algorithm to synthesize sensor data products. Figure~\ref{fig:viewingGeometry} illustrates the viewing geometry for ray tracing.  The sensor coordinate system is aligned to $ {u}-{v}-{w}$ reference frame with the origin located at $\mathbf{e}$, and the negative $\mathit{w}$ axis oriented towards the image plane, $\mathbf{P}$. The objects and light sources, located in the 3D virtual environment, are expressed in ${x}-{y}-{z}$ reference frame with origin $\mathbf{o}$. The ray tracing methodology involves casting a {ray} from the origin $\mathbf{e}$ of the sensor through each pixel in the image plane.

The intersection of the ray with any object in the scene is evaluated to assign a color to the pixels based on the intersection point. Note that each element of the rendering process, i.e., the ray and the 3D environment (comprising primitive geometries) are mathematical objects. Therefore, the intersection of the ray with these primitive geometries can be computed exactly. 
The mathematical expressions for evaluating ray-triangle and ray-sphere intersections adopted for this ray tracer engine are well-known in the literature~\citep{10.5555/1628957} and are also presented in~\ref{sec:rayObjIntersection} for completeness.
Further, owing to their co-planar nature and efficient software realization, triangular meshes are  adopted as primitive representations that build 3D geometric surfaces for models involving space based observations.

Using these ray-primitive intersections, the equation for shading the ray-surface intersections is formulated in the following section, along with methodologies to solve this equation.

\section{Mathematical Formulation of the Rendering Equation} \label{sec:s3_RenderingEq}

The interaction of light with the matter at the ray-surface intersection determines the \textit{shading} i.e., the color of the objects in the scene. The nature of shading depends on the optical properties and orientation of the objects and their surfaces, the amount of light (\textit{radiance}) arriving at a point in the scene, and the viewpoint perspective of the sensor or eye observing the scene. The interaction between the light and the surface is described using a multidimensional function known as \textit{Bidirectional Reflectance Distribution Function} (BRDF)~\citep{walter2007microfacet}, $\mathbf{f}_r$. The BRDF models the relationship between the incident radiance $L_i(\mathbf{x}, \hat{\boldsymbol{\omega}}_i)$ and the reflected radiance $L_r(\mathbf{x}, \hat{\boldsymbol{\omega}}_i, \hat{\boldsymbol{\omega}}_o)$ at a point $\mathbf{x}$ on an illuminated surface~\citep{shell2004bidirectional}. With $\hat{\boldsymbol{\omega}}_i$ describing the direction of incoming light and $\hat{\boldsymbol{\omega}}_o$, the direction of outing light ray, the BRDF can be expressed as 
\begin{equation} \label{eq:BRDF}
    \mathbf{f}_r(\mathbf{x},\hat{\boldsymbol{\omega}}_i,  \hat{\boldsymbol{\omega}}_o) = \frac{dL_r(\mathbf{x}, \hat{\boldsymbol{\omega}}_i, \hat{\boldsymbol{\omega}}_o)}{dL_i(\mathbf{x}, \hat{\boldsymbol{\omega}}_i)}
\end{equation}

\noindent Using the BRDF function in Eq. (\ref{eq:BRDF}), the reflected radiance at a point $\mathbf{x}$, with surface normal $\hat{\mathbf{n}}$, is computed by integrating all the incidence radiance arriving over a unit hemispherical area $\Omega$ as 
\begin{equation} \label{eq:10}
    L_r(\mathbf{x}, \hat{\boldsymbol{\omega}}_i, \hat{\boldsymbol{\omega}}_o) = \int_{\Omega} \mathbf{f}_r(\mathbf{x}, \hat{\boldsymbol{\omega}}_i,  \hat{\boldsymbol{\omega}}_o) L_i(\mathbf{x}, \hat{\boldsymbol{\omega}}_i) (\hat{\boldsymbol{\omega}}_i . \hat{\mathbf{n}}) d\hat{\boldsymbol{\omega}}_i
\end{equation}

\noindent Eq. (\ref{eq:10}) is a \textit{local illumination model}, and it represents only the local reflection of light as it hits the surface of the objects and does not capture absorption or refraction properties of the surface material. Hence, to truly capture the light-matter interaction, the total radiance leaving a surface,  $L_o(\mathbf{x}, \hat{\boldsymbol{\omega}}_o)$, is modeled as the sum of the radiance emitted by the underlying surface  $L_e $, and the reflected radiance,  $L_r$ 
\begin{equation} \label{eq:11}
    L_o(\mathbf{x},  \hat{\boldsymbol{\omega}}_o) = L_e(\mathbf{x}, \hat{\boldsymbol{\omega}}_o) + L_r(\mathbf{x}, \hat{\boldsymbol{\omega}}_i, \hat{\boldsymbol{\omega}}_o)
\end{equation}

\noindent Combining the Eqs. (\ref{eq:10}) and (\ref{eq:11}), we obtain the light-energy equilibrium approximated about a hemispherical differential area $\Omega$ at a surface point $\mathbf{x}$ as
\begin{equation} \label{renderingEq}
     L_o(\mathbf{x},  \hat{\boldsymbol{\omega}}_o) = L_e(\mathbf{x}, \hat{\boldsymbol{\omega}}_o) + \int_{\Omega} \mathbf{f}_r(\mathbf{x},  \hat{\boldsymbol{\omega}}_i, \hat{\boldsymbol{\omega}}_o) L_i(\mathbf{x}, \hat{\boldsymbol{\omega}}_i) (\hat{\boldsymbol{\omega}}_i . \hat{\mathbf{n}}) d\hat{\boldsymbol{\omega}}_i
\end{equation}

\noindent Eq. (\ref{renderingEq}) is referred to as the \textit{light transport equation} or the \textit{rendering equation}~\citep{kajiya1986rendering}. The solution to this equation determines the intensity of light at the surface point due to illumination from light  or other surfaces in the 3D scene. The intensities at all the other surfaces are computed similarly, adding to the recursive nature of the light energy distribution. The recursive formulation of the rendering equation suggests that the light emitters are not the only sources of scene illumination, but surfaces in the scene scatter and reflect light back into the world—making it a global illumination problem. The physically based ray tracer model described in this paper solves the global illumination problem to capture the reflection, refraction, and absorption effects of light at each of the ray-surface intersection points. 

The rendering equation (Eq. (\ref{renderingEq})) also calls upon the ray tracer to intelligibly follow or \textit{trace} the path of a primary light ray that is cast from a light source. The tracing of paths for all such secondary rays is quintessential for rendering realistic scenes to produce a cumulative \textit{global illumination} effect, and the ray tracer engine may also be addressed as a \textit{path tracer}. The ray tracer engine developed in this work implements random walk based Monte Carlo techniques for solving the rendering equation. The Monte Carlo solution converts the integral problem into an expected value problem with a path probability density function (PDF) $p(\Bar{x})$ for a sample path $\Bar{x}$ 
\begin{equation} \label{eq:13}
    I = \int f(\Bar{x}) dx = \int \frac{f(\Bar{x})}{p(\Bar{x})} . p(\Bar{x}) d\Bar{x} = E\left[\frac{f(\Bar{x})}{p(\Bar{x})}\right]
\end{equation}

The expected value is estimated from $N$ random sample paths $\Bar{x}_1, \Bar{x}_2, \Bar{x}_3, ..., \Bar{x}_N$ generated with the probability density function $p$ 
\begin{equation} \label{eq:14}
    I =  E\left[\frac{f(\Bar{x})}{p(\Bar{x})}\right] \approx \frac{1}{N} \sum_{i = 1}^{N} \frac{f(\Bar{x}_i)}{p(\Bar{x}_i)}
\end{equation}

\begin{figure}[ht]
\centering
\subfloat[Multiple rays through each pixel to obtain average pixel intensity.]{\label{fig:pixelSampling} \includegraphics[width=0.4\textwidth]{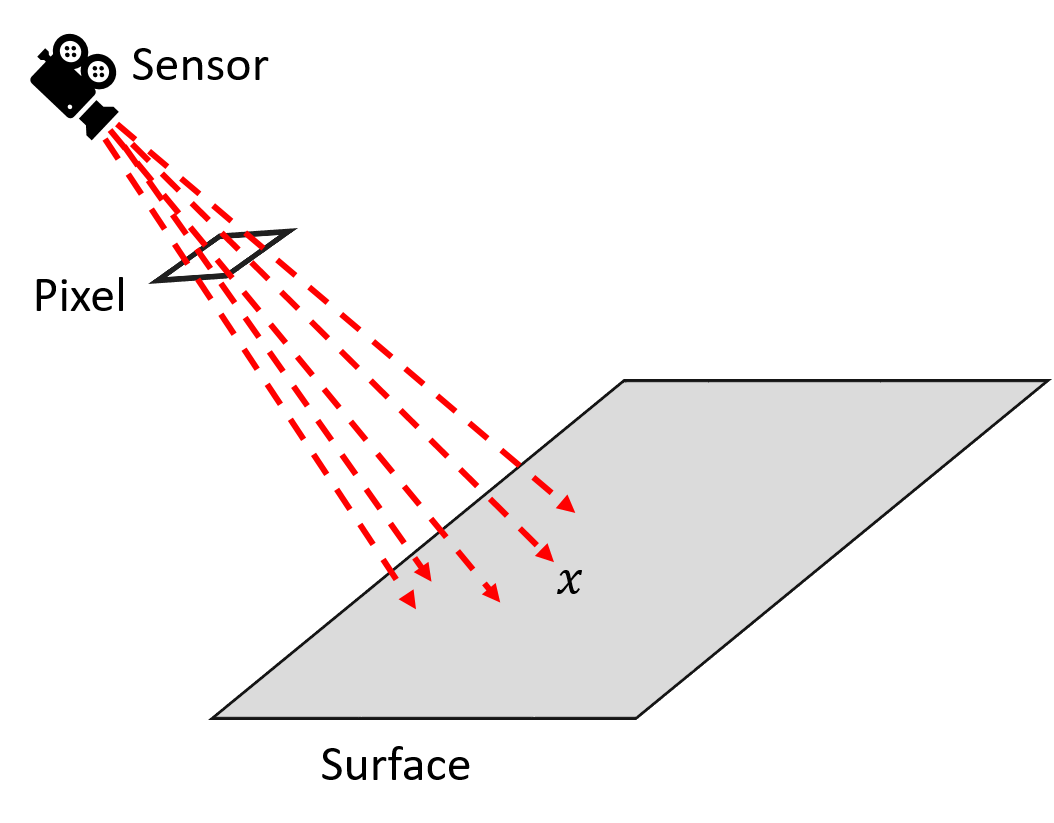}}\hfill
\subfloat[Tracing path from sensor to the light source.] {\label{fig:pathTracing} \includegraphics[width=0.4\textwidth]{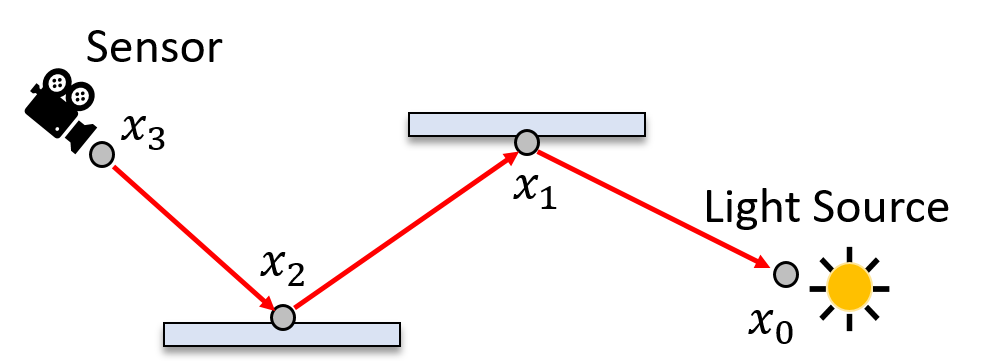}} 
\caption{Pixel sampling in ray tracing.}
\label{fig:samplingRT}
\end{figure}

\comment{
\begin{figure}[ht]
        \centering
        \caption{Multiple rays through each pixel to obtain average pixel intensity.}
        \comment{
        \caption{This figure demonstrates the sampling of the ray tracer engine. The ray tracer shoots multiple rays through each pixel, in the directions specified by the BRDF. The intensities obtained in all the sample directions are averaged to evaluate the corresponding pixel intensity.}
        }
         \label{fig:pixelSampling}
        \end{figure}

\begin{figure}[!t]
        \centering
        \caption{Tracing path from sensor to the light source.}
        \comment{
        \caption{The ray tracer engine traces the path from the sensor to the light source to approximate the illumination at each surface point from the reflections at all other possible surfaces in the scene.}}
         \label{fig:pathTracing}
        \end{figure}
}

Equation (\ref{eq:14}) indicates that the contribution from all the samples is averaged to estimate the pixel intensities. This phenomenon is illustrated in Fig. \ref{fig:pixelSampling}. \cite{cook1984distributed} first presented this idea of stochastic or random sampling under the banner of \textit{distributed ray tracing}. Rather than uniform sampling in all directions, Monte Carlo integration benefits from taking samples in the random directions proportional to the BRDF. Basing the probability distribution of the samples on the BRDF reduces the likelihood of choosing to sample directions near the horizon, where there is miniscule contribution to the illumination. This choice of sampling strategy, also called \textit{importance sampling} reduces the variance because the estimator is close to the shape of the actual function. The importance sampling based unbiased estimation defines the PDF for a full path $\Bar{x}$ as a product of conditional PDFs at the $n$ path vertices $x_0, \, x_1, \, ... x_n$ as shown in Eq. (\ref{eq:15}) and illustrated in Fig.~\ref{fig:pathTracing}, where the full path is traced from the sensor to the light source (or sources) in order to approximate the illumination via Monte Carlo integration. 
\begin{equation} \label{eq:15}
    p(\Bar{x}) = p(x_0,x_1,x_2,x_3) = p(x_3)p(x_2|x_3)p(x_1|x_2)p(x_0)
\end{equation}
This approximation of the pixel intensities improves with increasing number of samples per pixel, which is essential for rendering realistic images void of aliasing visual artifacts. 





\subsection{Lighting model}

While the evaluation of path tracing integral approximates pixel illumination, the optical properties that contribute to the illumination at a surface are governed by its material properties. BRDFs model the behavior of the material reflectance, at each surface points, for every incoming and outgoing directions as represented in Eq. (\ref{eq:BRDF}). Fundamentally, the BRDF is a measure of the amount of light scattered by a surface from one direction to another. Integrating the BRDF in Eq. (\ref{eq:10}) over a specified incident and reflected solid angles determines the reflectance at a surface point. When a beam of light strikes a surface interface, it either gets scattered at the top or might penetrate into the material and undergo sub-surface or volumetric scattering. The majority of the scattering phenomenon occurs at the top layer, and it primarily depends on the refractive index of the illuminated medium. The topography of the interface determines the angular distribution of the scattered radiance  - smooth surfaces reflect light into a specular direction, while rough surfaces reflect and diffuse light in many directions. Ideal diffusely reflecting or \textit{Lambertian} objects appear equally bright from all viewing directions. Ultimately, the information on the nature of light-surface interaction is determined by the BRDF models for the objects. The BRDF models can be obtained from empirical measurements for various materials, which are fit into a mathematical function. Most BRDF models are generally a combination of specular and diffuse components. The Lambertian BRDF model assumes a constant BRDF value of, $\frac{\rho}{\pi}$ where $\rho$ (albedo) is the measure of diffuse reflection. While the Lambertian BRDF is effective for diffuse body reflections, it cannot accurately model specular surface reflections. The Phong lighting model~\citep{phong1975illumination} captures both diffuse reflections from rough surfaces and specular reflections from smooth surfaces. In this work, the Phong lighting model is considered to approximate the BRDF of space-object surfaces. The equation for the Phong model to compute the illumination at a surface point $x_p$ with normal $\hat{\mathbf{N}}$, per unit area perpendicular to the viewing direction $\hat{\mathbf{V}}$, is a combination of emissive ($I_E$), ambient ($I_A$), diffuse ($I_D$), and specular ($I_S$) illumination intensities.  
\begin{equation} \label{eq:phongequation}
    I(x_p, \hat{\mathbf{N}}, \hat{\mathbf{V}}) = k_e I_{E} + k_a I_{A} + k_d I_{D}(\hat{\mathbf{L}}. \hat{\mathbf{N}}) + k_s I_S (\hat{\mathbf{R}} \cdot \hat{\mathbf{V}})^n
\end{equation}
here, 
$k_e$, $k_a$, $k_d$, and $k_s$ are emissive, ambient, diffuse and specular constants that are exclusive to the type of surface materials. $\hat{\mathbf{L}}$ is the direction of light incidence and $\hat{\mathbf{R}}$ is the direction of reflection. The constant $n$ is used to regulate the size of specular highlights based on the material properties.

\begin{figure}[t]
  \centering
  \subfloat[Ideal diffuse, or Lambertian BRDF reflecting light equally in all directions]{\includegraphics[width=0.3\textwidth]{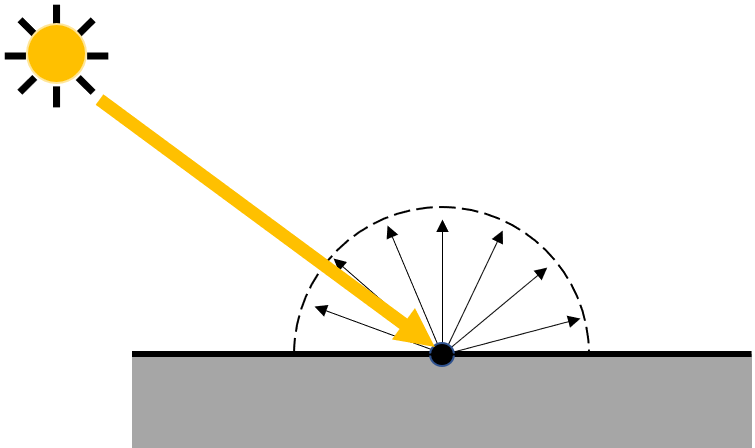}{\label{fig:Lambertian}}}
  \hfill
  \subfloat[Specular material reflecting light in the mirror direction]{\includegraphics[width=0.3\textwidth]{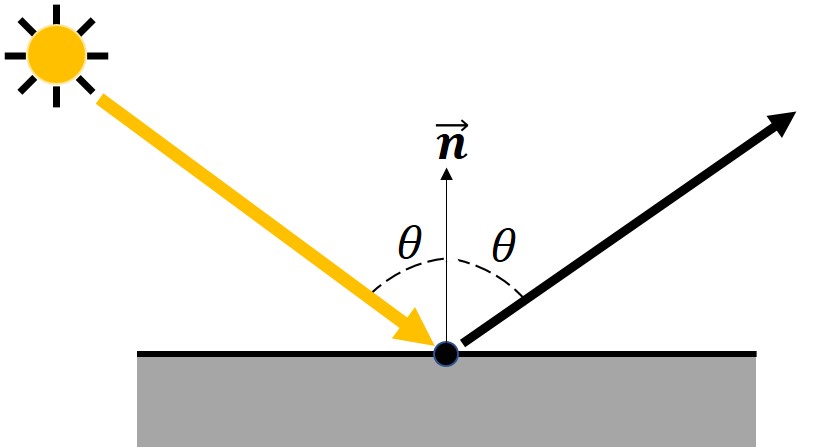}{\label{fig:specular}}}
  \caption{Lambertian and specular surfaces.}
  \label{fig:LambertianSpecular}
\end{figure}

Recall that a ray cast into the scene might not stop upon hitting an object, and may undergo reflection or refraction at the ray-surface intersection and further cast into the scene. The ray tracer based rendering program under discussion is a global illumination method that accounts for shadows and effects such as reflection, refraction, scattering, indirect reflection, and indirect diffuse. The effect of shadows is determined by computing if a ray from a surface point hits an object on its path to the light sources.  If it does, the point is categorized to be in a shadow, as shown in Fig. \ref{fig:shadows}. In the demonstration shown, the surface point $\mathbf{p}$ is in shadow while the point $\mathbf{q}$ is not in shadow.

\begin{figure}[!t]
        \centering
        \includegraphics[width=0.3\textwidth]{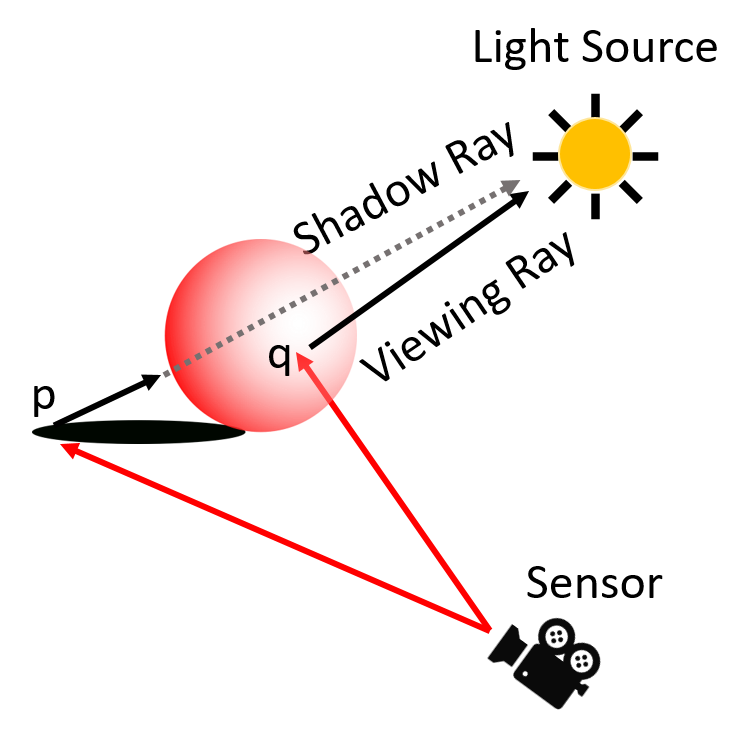}
        \caption{Computing shadows by tracing the light from the sensor to the source.}
         \label{fig:shadows}
        \end{figure}

The specular reflection (reflection of light in mirror direction) is evaluated using the specular term of the Phong lighting model described in Eq. (\ref{eq:phongequation}). The reflected ray $\hat{\mathbf{R}}$ \comment{in Eq. (\ref{eq:phongequation})} is computed according to the law of reflection as
\begin{equation}
    \hat{\mathbf{R}} = \hat{\mathbf{V}} + 2 \left(\hat{\mathbf{V}}.\hat{\mathbf{N}} \right) \hat{\mathbf{N}}
\end{equation}
\noindent The rendering algorithm evaluates the reflection for a maximum recursion depth of 40 to balance the speed of the rendering process and the accuracy of the reflection effect. Snell's law of refraction is implemented to model the effects of bending of light at a specular interface, between two media with refractive indices $n$ and $n_t$, $\theta$ angle of incidence and $\phi$ angle of refraction, as
\begin{equation}
    n \sin \theta = n_t \sin \phi
\end{equation}

\noindent The transmitted ray $\hat{\mathbf{T}}$ can be described from the basis formed by the orthonormal vectors $\hat{\mathbf{N}}$ and $\hat{\mathbf{B}}$ as
\begin{equation}
    \hat{\mathbf{T}} = \sin \phi \hat{\mathbf{B}} - \cos \phi \hat{\mathbf{N}}
\end{equation}
\noindent $\hat{\mathbf{B}}$ can also be obtained from the $\hat{V}$ described in the same basis,
\begin{align}
    \hat{\mathbf{V}} &= \sin \theta \hat{\mathbf{B}} - \cos \theta \hat{\mathbf{N}} \\
    \hat{\mathbf{B}} &= \frac{\hat{\mathbf{V}} + \hat{\mathbf{N}} \cos  \theta}{\sin \theta}
\end{align}

The reflection of light from a non-conducting or dielectric interface between two media varies according to the Fresnel equations. Schlick's approximation~\citep{schlick1994inexpensive} method approximates the Fresnel equations for an efficient computation of reflectivity ($R$) as
\begin{equation}
    R(\theta) = R_0 + (1 - R_0) (1-\cos \theta)^5
\end{equation}

\noindent where $R_0$ is the reflectance at normal incidence: 
\begin{equation}
    R_0 = \left (\frac{n_t - 1}{n_t + 1} \right)^2
\end{equation}

Furthermore, for planetary photometry, custom BRDF models such as the Lommel-Seeliger~\citep{hicks2011photometric}, Hapke~\citep{hapke1984bidirectional}, or the Shkuratov~\citep{shkuratov2011optical} can be implemented using the NaRPA's C++ software framework. 

\subsection{Acceleration structure of the ray tracer engine}
\noindent The recursive evaluation of rendering equation (Eq. (\ref{renderingEq})) evaluates each ray against every intersection in the scene. This brute force search for ray-surface intersection is inefficient and computationally very expensive. Kay and Kajiya~\citep{kay1986ray} presented a technique called \textit{bounding volume hierarchy} (BVH) as a method to accelerate ray tracing. The fundamental idea of BVH is to place multiple bounding boxes around all the objects in the scene geometry. The bounding boxes are hierarchically-arranged as a tree-based structure, with a large bounding box as a root and smaller bounding boxes within it as subtrees. This is illustrated in Fig.~\ref{fig:bvh}. As shown, bounding boxes created around the geometric primitives of the model and are nested inside another bounding box. Using the BVH technique, it is tested if a ray intersects with the volume defined by the bounding boxes instead of the entire scene space to speed up the computation of ray-surface intersection. The engine creates a BVH structure from the source geometry prior to rendering a scene.
\begin{figure}[t]
        \centering
        \includegraphics[width=0.45\textwidth]{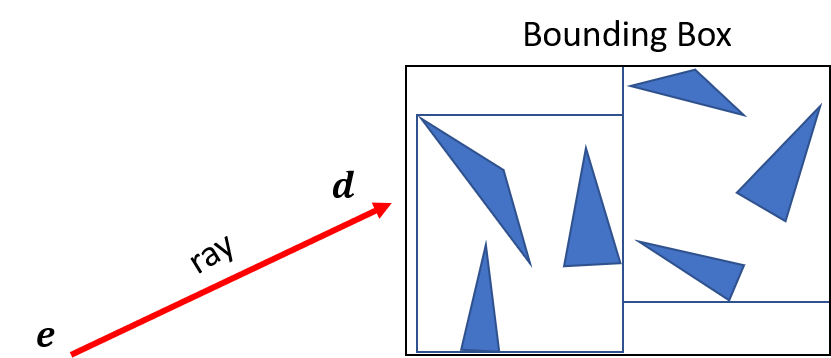}
        \caption{Bounding volume hierarchy.}
         \label{fig:bvh}
        \end{figure}

\subsection{Capabilities} \label{sec:rayTracerCap}

The capabilities of the rendering engine NaRPA are outlined in the rest of this section. These capabilities include texture mapping, global illumination effect, depth of field effect, point cloud, depth map, and contour map generation. They are described as follows:  

\begin{itemize}[leftmargin=*]
    \item Texture mapping: 
    Texture mapping \comment{\cite{catmull1974subdivision}} technique allows one to project a complex pattern (\textit{texture}) on a simple polygon. The pattern may be repeated many times on to tile a plane, or it can be a multidimensional image projected onto a multidimensional space. 2D texture mapping associates the 2D coordinates $(u,v) \in [0,1]^2$ that correspond to a location in the texture image with points on a 3D surface. This is accomplished in two phases. First the $(u,v)$ space is mapped to object space $(x_o, y_o, z_o)$ by parameterizing the surface followed by transforming the object onto a screen, by the rendering engine. As an example, spherical coordinate parameterization to map a texture image onto a sphere is shown in Fig.~\ref{fig:textureMap}. With the renderer's ability to generate texture mapped primitives, clusters of objects can be conveniently represented to maintain high frame rates and high level of details to aid in visual navigation ~\citep{maciel1995visual}.
    \begin{figure}[t]
  \centering
  \subfloat[Earth surface texture map as a 2D image]{\includegraphics[width=0.45\textwidth]{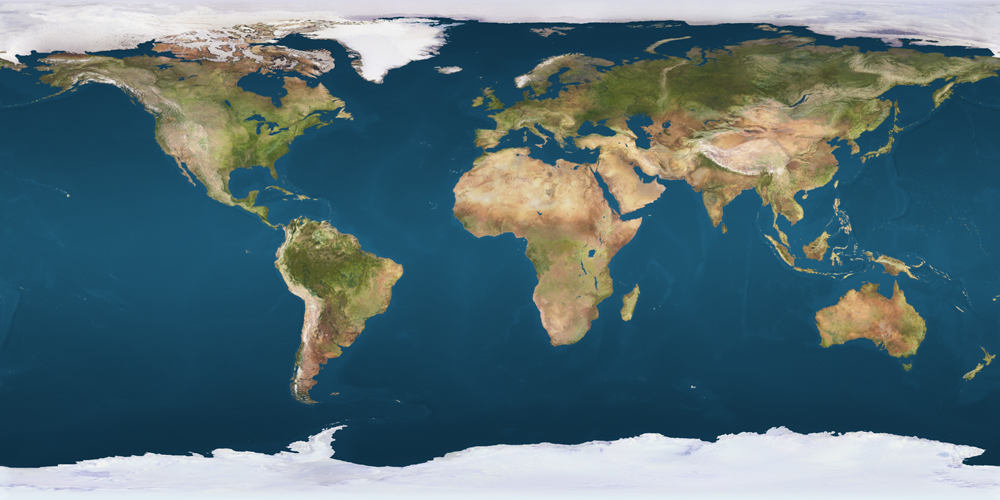}{\label{fig:earthMap}}}
  \hfill
  \subfloat[Texture projected onto a sphere]{\includegraphics[width=0.25\textwidth]{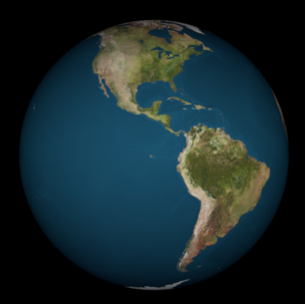}{\label{fig:earthProjection}}}
  \caption{3D texture mapping capabilities.}
  \comment{
  \caption{This figure illustrates the 3D texture mapping capabilities of the ray tracer. 2D image coordinates that correspond to the Earth are projected onto a sphere to model the Earth in free space.}}
  \label{fig:textureMap}
\end{figure}
    
Let $-\pi \leq \theta \leq \pi$ be the latitude angle around the sphere in the horizontal plane and $0 \leq \phi \leq \pi$ be the longitude angle measured from $0$ at the South Pole to $\pi$ at the North Pole. The parametric equation of a sphere with radius $R$ and center $(x_c, y_c, z_c)$ is
\begin{align}
        x_o &= x_c +  R \cos \phi \sin \theta
        \\
        y_o &= y_c + R \sin \phi \sin \theta
        \\
        z_o &= z_c + R \cos \theta
\end{align}
    and 
\begin{align}
        \theta &= \cos^{-1} \left(\frac{z_o - z_c}{R} \right)
        \\
        \phi &= \tan^{-1} \left(\frac{y_o - y_c}{x_o - x_c} \right)
\end{align}
Now, defining 
    \begin{align}
        u &= \frac{\phi}{2\pi} \\
        v &= \frac{\pi - \theta}{\pi}
    \end{align}
    
    allows us to index the $(u,v)$ of the texture map for all the points on the surface of the sphere. This mapping is shown in the Fig. \ref{fig:textureMap}.

    \item Global illumination: 
    The ray tracing engine presented in this paper is a global illumination algorithm. Simulating the effects of both direct and indirect illuminations to render a realistic image from the scene description defines the term global illumination. Direct illumination is a result of intersection between light rays from the source and the object surface. Indirect illumination stems from the multiple complex reflections that the rays undergo before exiting the scene geometry. 
    This effect is a key aspect to generate reliable simulations for developing computer vision algorithms aimed at vision-based navigation~\citep{lebreton2021image}.  
    

    \item Depth of field effect: 
    Imaging sensors have an aperture through which light enters the sensor to register an image on the sensor film. A smaller aperture increases the depth of the scene which is in focus, and the objects in the scene appear sharp when rendered. Using a larger aperture, in combination with a lens, allows more light into the sensor but is less controlled. The 3D objects at the focal length are projected into a single point on the sensor, while the other points map into a blur circle. The latter effect, resulting in a blurred portion of the image from objects at certain distance, is known as \textit{depth of field} or the \textit{defocus blur} effect. Defocus effects are commonly encountered because of large relative distances among planetary objects. Therefore, it is essential to simulate navigation scenarios with attribution to the defocus and motion blur effects\comment{ \cite{harris2021visual}}.
\comment{
    \begin{figure}[t]
        \centering
        \caption{Thin lens system with the light rays from the object converge at a point on the view plane, reckoning the object point to be in focus.}
         \label{fig:lens}
        \end{figure}
    }    
    The ray tracing engine adopts a thin lens model for the virtual camera to generate depth of field effects~\citep{potmesil1981lens}. In the thin lens model, the rays of light emitted from a surface point travel through the lens and converge at a point behind the lens. The focal length (distance between the point of projection and the image plane), $f$, defines this characteristic of the thin lens and is modeled using the lens equation
    \begin{equation} \label{lensequation}
        \frac{1}{f} = \frac{1}{||z_0||} + \frac{1}{||z_1||}
    \end{equation}
    
The thin lens approximation for the virtual camera of the ray tracer engine simulates the depth of field effect by controlling the \textit{focus distance} (distance from the point of projection and the object plane in focus) and/or the aperture radius. From Eq. (\ref{lensequation}), the distance between the view plane and the lens (projection point) for a given focal length $f$ and a focus distance $d_f$ is
    \begin{equation} \label{eq:thinLens}
        z_0 = \frac{fd_f}{d_f - f}
    \end{equation}
Eq. (\ref{eq:thinLens}) represents the distance at which the projection of an object or a point that is in focus at a distance $d_f$ in front of the lens, will converge behind the lens. If the film is located at $z_0$, the projection of the object converges to a point and otherwise to a circle of confusion (COC) to appear as a blurred spot. Figure~\ref{fig:depthOfField} shows the depth of field effect with a fixed focus distance and increasing aperture radius to observe the defocus blur effect on the ISS with the moon in the background.  
\begin{figure*}[t]
\centering
\subfloat[Aperture radius: 0]{\label{fig:a}\includegraphics[width=0.2\textwidth]{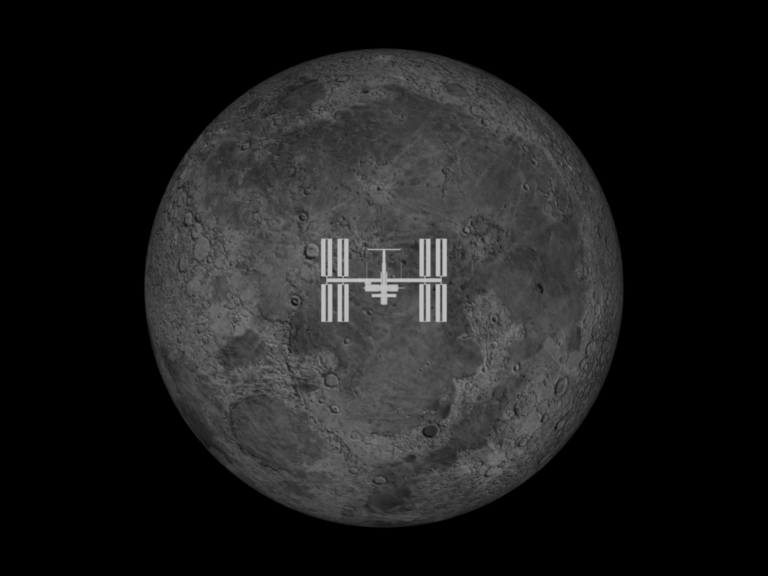}}\quad
\subfloat[Aperture radius: 0.04]{\label{fig:b}\includegraphics[width=0.2\textwidth]{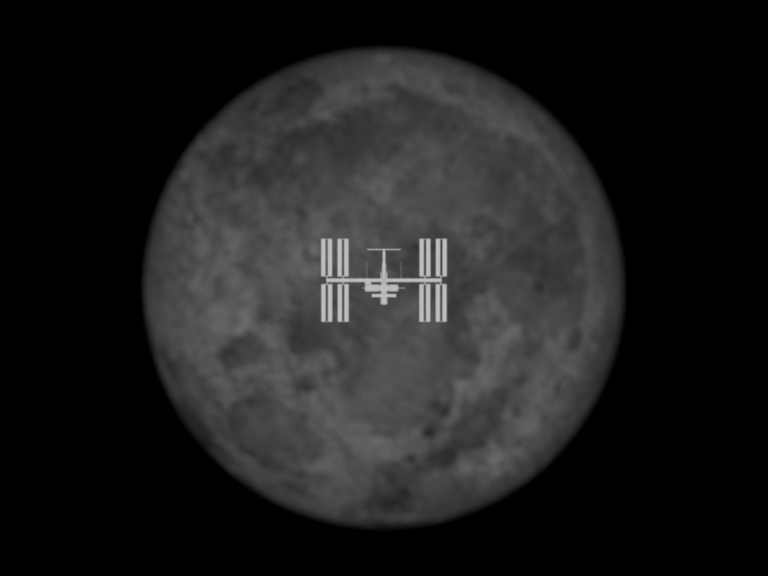}}\quad
\subfloat[Aperture radius: 0.06]{\label{fig:c}\includegraphics[width=0.2\textwidth]{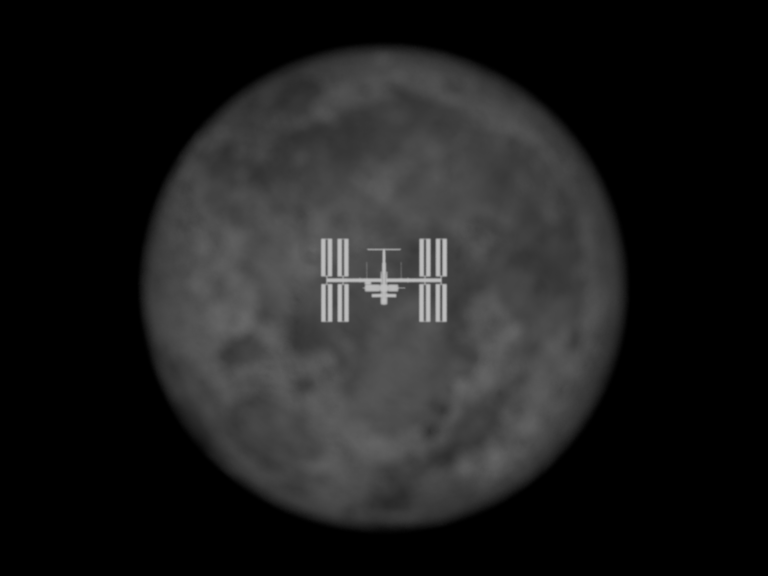}}\quad%
\subfloat[Aperture radius: 0.15]{\label{fig:d}\includegraphics[width=0.2\textwidth]{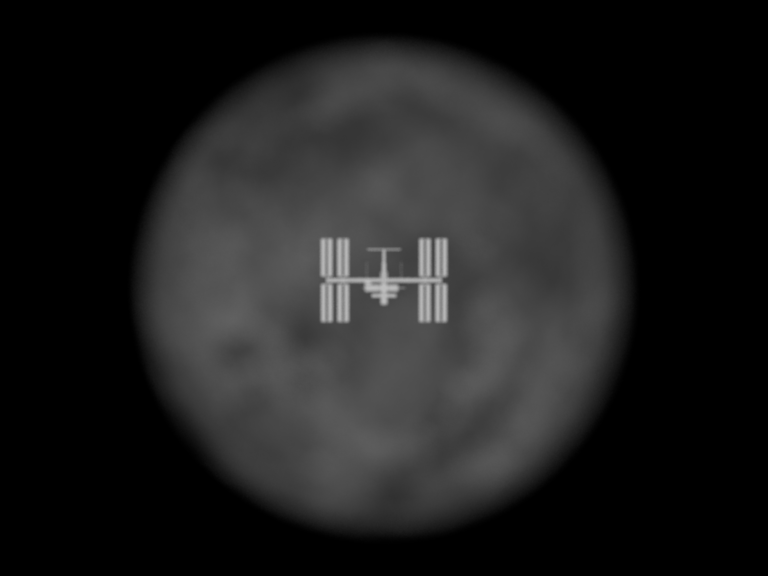}}%
\caption{Defocus blur effect with incremental blurring effect due to widening of the camera aperture.}
\comment{
\caption{This figure illustrates the defocus blur effect with the International Space Station (ISS) in focus and an incremental blurring effect on the moon due to widening of the camera aperture (scene units).}
}
\label{fig:depthOfField}
\end{figure*}

 \begin{figure*}[t]
\centering
\subfloat[3D point cloud data of Curiosity landing site]{\includegraphics[width=0.4\textwidth]{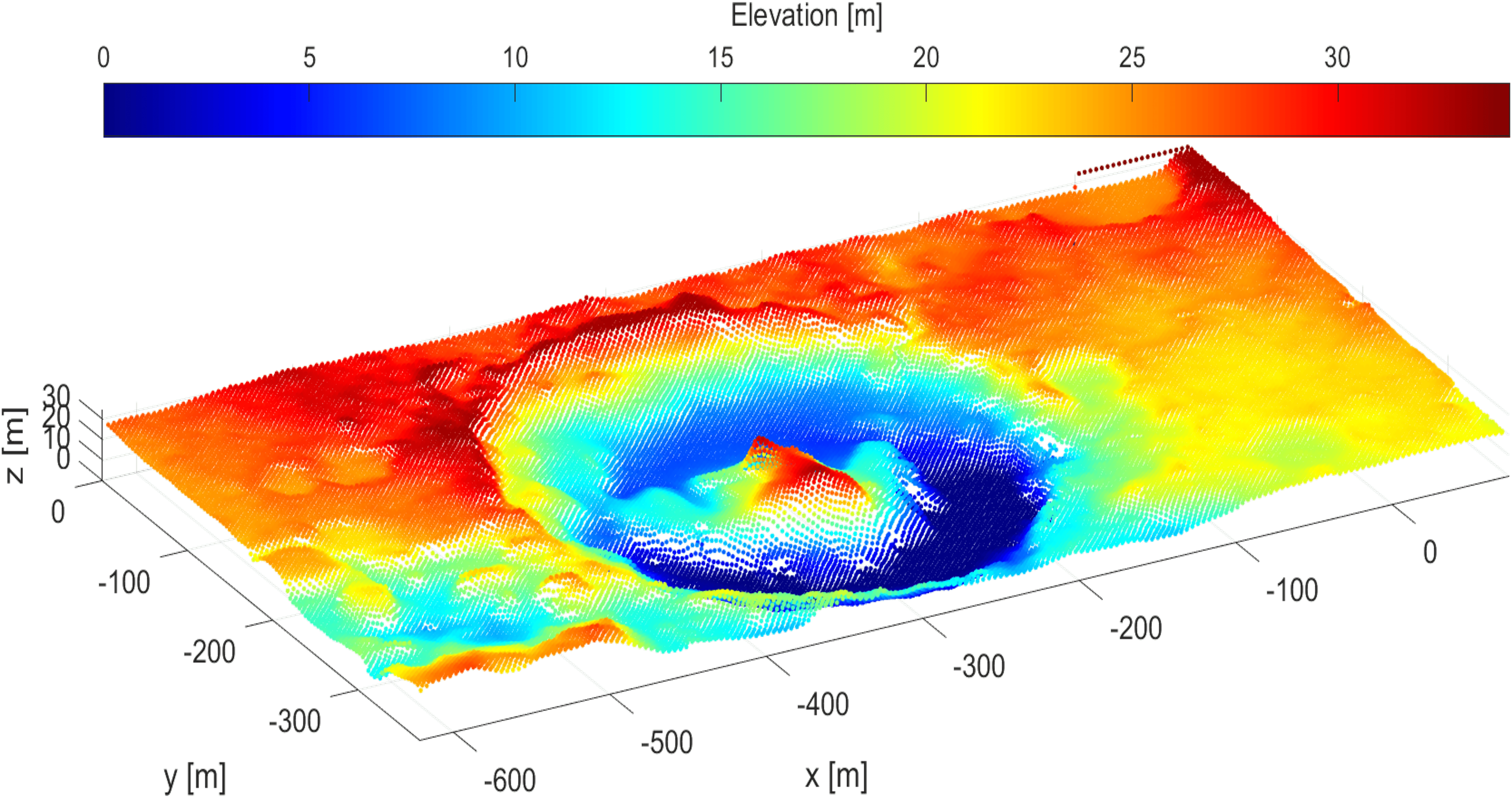}{\label{fig:pointCloud}}}
\quad
\subfloat[Depth map of the Curiosity landing site]{\includegraphics[width=0.23\textwidth]{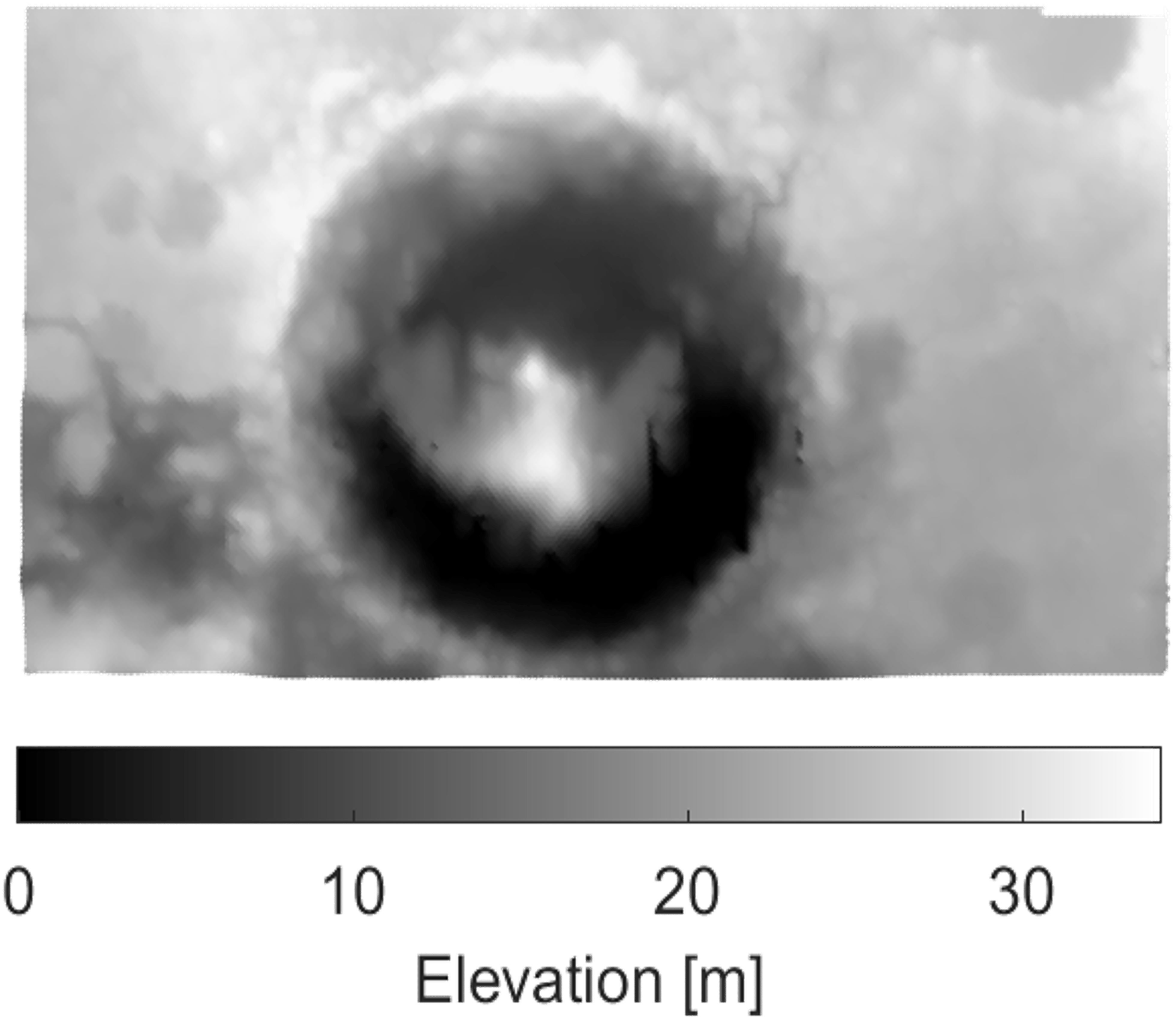}{\label{fig:depthMap}}}
\quad
\subfloat[Contour map of the Curiosity landing site]{\includegraphics[width=0.25\textwidth]{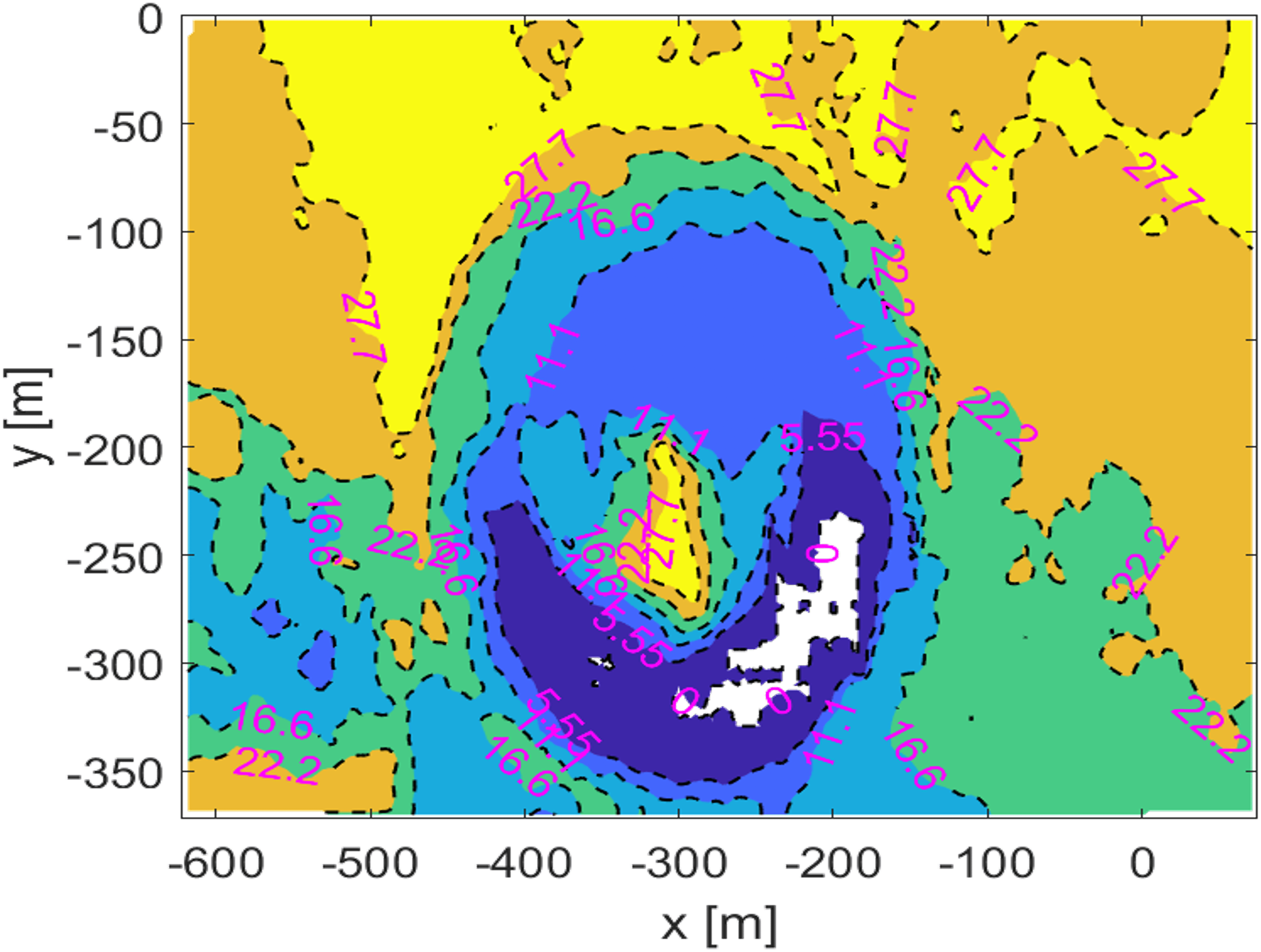}{\label{fig:contourMap}}}
\caption{Point cloud, depth, and contour map capabilities.}
\comment{
\caption{This figure demonstrates the simulation of a LiDAR sensor configured on a spacecraft approaching vertically down. The data corresponding to point cloud, depth and contour maps is obtained from the ray tracer engine configured with the virtual LiDAR scanner.}
}
\label{fig:PointCloud_DMap}
\end{figure*}  

\item 3D Point Cloud, Depth, and Contour Maps: 
    \comment{3D point clouds are a collection of 3D points located in space and are associated with a color.} Point cloud data is generally acquired by 3D imaging sensors such as LiDAR (Light detection and ranging). The rendering software stores the 3D location of the first point of intersection between each ray and the objects in the scene. It uses the light propagation time to compute range to different objects in the scene. This simulates the characteristic feature of LiDAR type scanners and depth cameras and thereby takes advantage of the existing ray tracer algorithm to generate point clouds and depth maps. As will be shown in section~\ref{sec:velocimeterLidar}, the NaRPA engine also accommodates simulation of velocimetry LiDAR sensing from continuous sweeps of LiDAR scans. Figure~\ref{fig:PointCloud_DMap} shows the point cloud data extracted from the digital elevation model (DEM) of Curiosity landing site within the Gale crater of Mars. Depth and contour maps are generated from the elevation data of the DEMs. By the inclusion of physically-based simulator for LiDAR sensors, ray-tracers  have the capabilities to generate true physical data and enable testing of registration algorithms for spacecraft relative navigation~\citep{christian2013survey, ramchander2021hardware}.  
    
 \comment{   Depth data is the distance $t$ from the virtual camera $e$ to the ray-primitive intersection point. }  

 The sensing and rendering capabilities are further described through the application of this graphical engine for remote sensing observations as will be described in the following sections.

\end{itemize}
\section{Atmospheric Modeling and Rendering Ground-based Observation of Space Objects} \label{sec:atmosphericModel}

The scattering of light by particles present in the atmosphere influences the appearance of an object. Particles or gas molecules redirect light rays from their original path and affect the colors of the object observed in natural light. The methods to simulate the scattering effects of the atmosphere are described by     ~\cite{nishita1985continuous}. ~\cite{10.5555/1628957} proposed an analytical model which provides an accurate simulation of the sky colors and is more restrictive in that the model only works for an observer located on the ground. The methods of atmospheric scattering simulations in this paper borrows from the work from~\cite{nishita1985continuous,10.5555/1628957} and is outlined in this section. Additionally, the contributions from~\cite{tessendorf1988comparison,tessendorf1994impact} have been adopted to embed multiple scattering effects for atmospheric rendering.   
 
 \subsection{Atmospheric model}
 Atmosphere is a thin blanket of gases that envelopes a planet. The atmosphere is held in place by the gravity of the planet and is defined in multiple layers on the basis of its temperature distribution and composition. The thickness of the Earth's atmosphere is about 100 km and is made of oxygen, nitrogen, argon, carbon dioxide, water vapor, and other floating particles classified as aerosols, including dust, pollen, smoke, and more. In this paper, a thickness of 60 km is used by considering scattering of light rays in only the two lower layers of the Earth's atmosphere, the troposphere and the stratosphere. Photons that travel through the Earth's atmosphere in a certain direction are deflected in various directions when they collide with the particles in the atmosphere. The dispersion and diffusion of light as a result of atmospheric scattering depends on the size and density of particles in the atmosphere. The density or tightness of these particles decreases with altitude. Thus, the number of molecules per cubic volume is relatively low in the upper layers of the atmosphere.
 
 The scattering is computed by sampling the atmospheric density along the light ray at regular intervals. As a first step, the scattering of light is approximated by a forward refraction~\citep{nishita1985continuous}. A fundamental approach is to determine the scattering from the assumption of an exponential decrease in the atmosphere density with altitude modeled with an equation of the form:
\begin{equation}\label{den1}
\rho(h) = \rho(h_0) \exp {\Big(-\frac{h}{H}\Big)}   
\end{equation}
\noindent where $\rho(h_0)$ is the density of air at sea level, $h$ is the current altitude above the sea level and $H$ is the thickness of the atmosphere if its density were uniform. Using this information on density and a similar model of the pressure distribution of air molecules in the atmosphere, we utilize a model that states the refractive index of the atmosphere at a particular height above the Earth's surface. Exponential refractive index model with altitude dependence of refractive index is given by 

\begin{align}
    &(n(h)-1)\times 10^6 = \frac{79 \left( P + \frac{4800e}{T} \right)}{T};
    \\
    &\begin{cases}
    \text{Troposphere ($<12$ km)}& T = 59 - 0.00356 h\\ & P = 2116 \left( \frac{T+459.7 }{518.6} \right)^{5.256} \\
    \text{Upper atmosphere ($>12$ km)}& T = -70\\ & P = 473.1 \times e^{( 1.73 - .000048 h)}
\end{cases}
\end{align}

Here, $n(h)$ is the average refractivity at height $h$, $T$ is the temperature, $P$ is the pressure and $e$ is the density of water vapor and other gases given by Eq. (\ref{den1}). The refractive index is modeled as a series of spherically concentric shells for computational advantage. The atmosphere modeled as a series of spherical shells considers a higher number of samples at lower altitudes where the density is high and a lower number of samples at higher altitudes where the air is less dense. This technique of `importance sampling' improves upon the convergence speed relative to a `constant step' integration method. Since the density increases exponentially with altitude, numerical integration involved in ray tracing is performed over small intervals at low altitude and larger intervals at high altitudes. This setup is illustrated in Fig.~\ref{F18}.

 \begin{figure}[ht]
    \centering
    \includegraphics[width=0.49\textwidth]{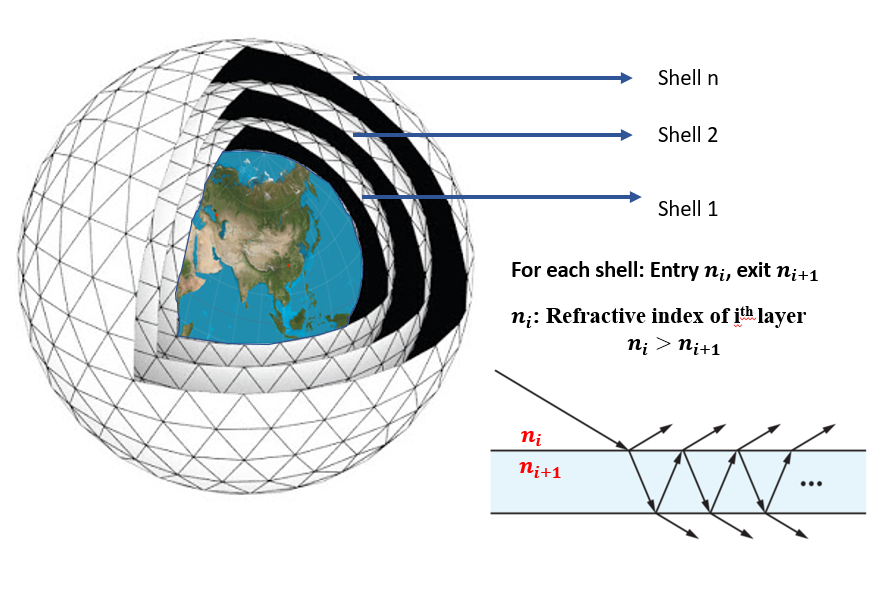}
    \caption{Refractive Index model.}
    \label{F18}
\end{figure}

The scattering of light depends strongly on the size of the particles in the atmosphere. If the scattering is due to particles that are smaller than the wavelength of the light, then it is called Rayleigh scattering. In Rayleigh scattering, incident light is scattered more heavily at the shorter wavelengths. If the incident light is evenly scattered in all directions, it is called Mie scattering. Larger particles in the air called aerosols (dust, pollen) cause Mie scattering. To capture the effect of Rayleigh scattering, we modeled the spherical shells with varying refractive indices for each sample of the ray generated and the depth for each ray to be slightly different (using a pseudorandom number). Since the aerosol distribution is only limited to the lower atmosphere, Mie scattering effects are captured using the proposed method for spherical shells lying below a particular altitude. 

Note here that in the description of our model the Sun is the source of illumination of the sky. The Sun is assumed to be at a considerable distance from Earth, and the light rays reaching the atmosphere are parallel to each other. This is similar to the sky emitter built into the Mitsuba renderer~\citep{jakob2013mitsuba}. 

\subsubsection{Atmospheric modeling and ground-based space object simulation results}
\begin{figure*}[ht!]
\centering
\subfloat[7:20 AM]{\label{fig:toda}\includegraphics[width=0.21\textwidth]{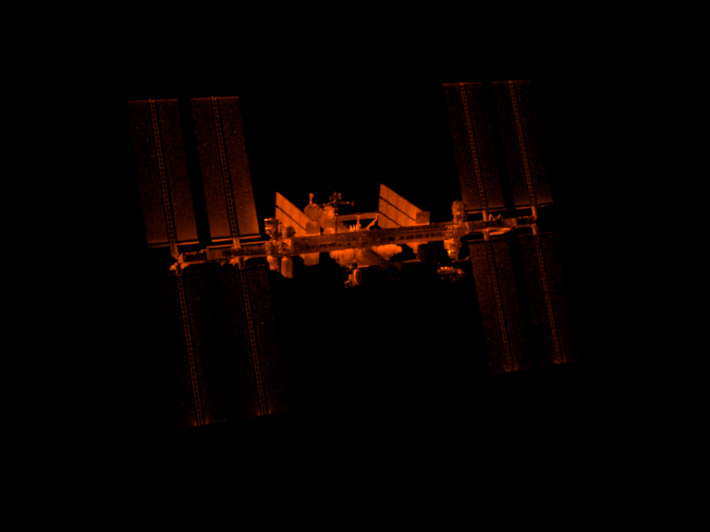}}\qquad
\subfloat[8:20 AM]{\label{fig:todb}\includegraphics[width=0.21\textwidth]{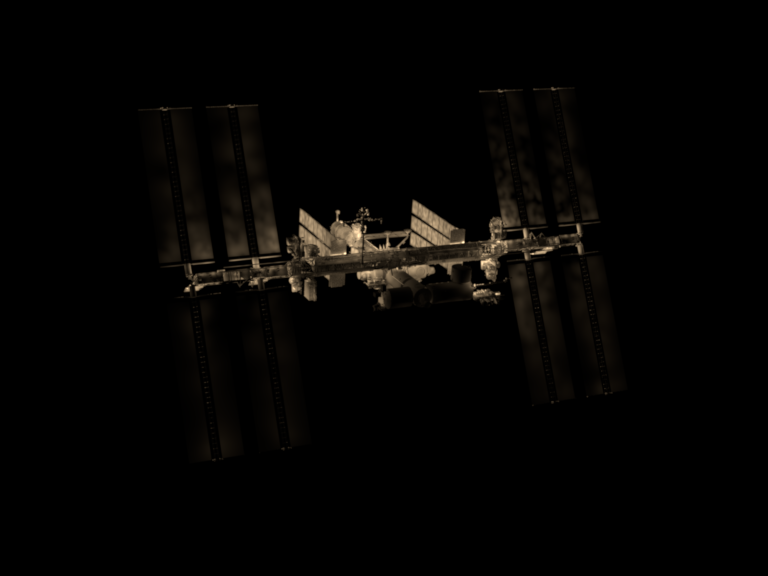}} \qquad
\subfloat[6:20 PM]{\label{fig:todc}\includegraphics[width=0.21\textwidth]{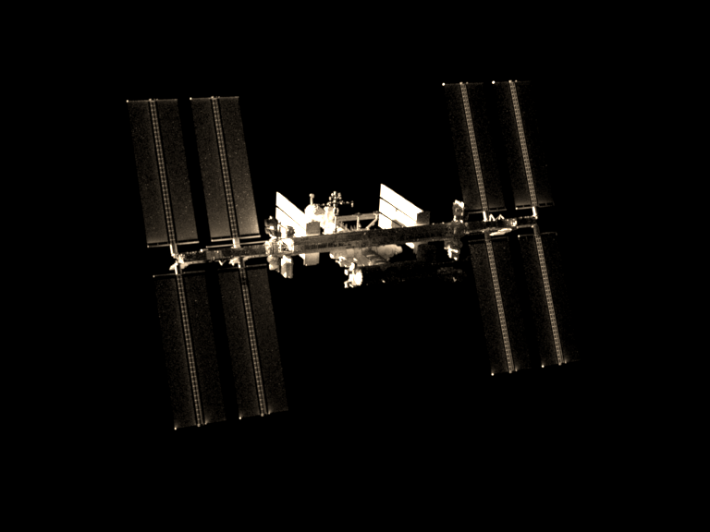}}\qquad%
\subfloat[7:20 PM]{\label{fig:todd}\includegraphics[width=0.21\textwidth]{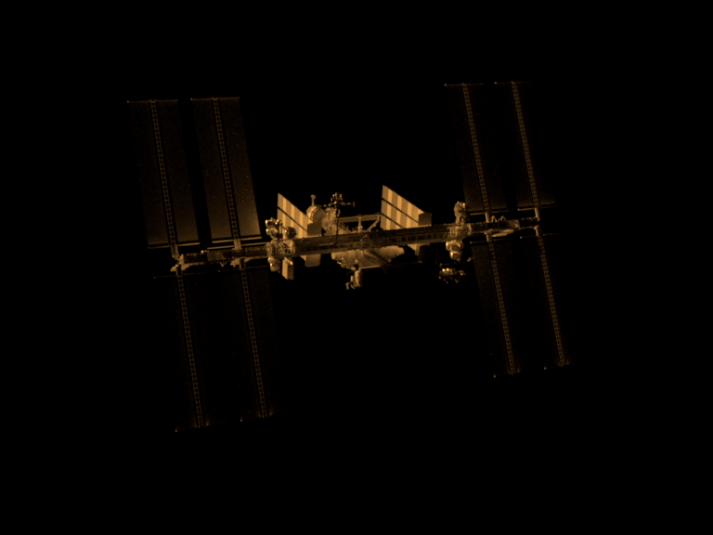}} 
\caption{Ground-based observation of the ISS, illustrating renderings at different times of the day.}
\label{fig:timesOfDay}
\end{figure*}

\begin{figure*}[ht!]
\centering
\subfloat[Turbidity level: 1]{\label{fig:ea}\includegraphics[width=0.21\textwidth]{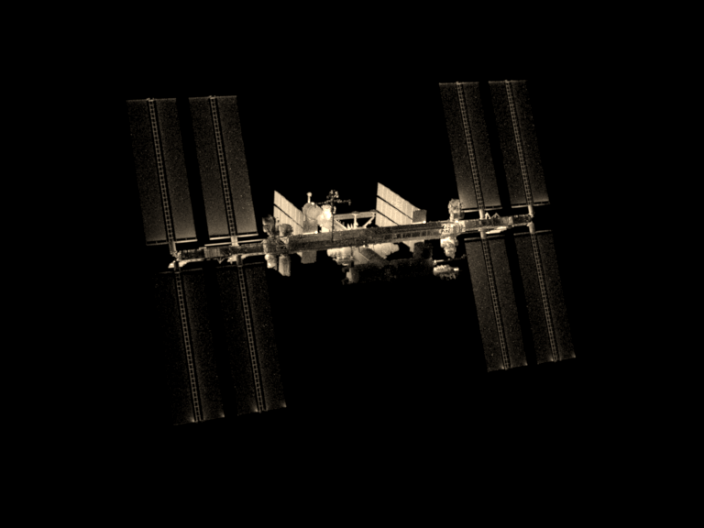}}\qquad
\subfloat[Turbidity level: 3]{\label{fig:eb}\includegraphics[width=0.21\textwidth]{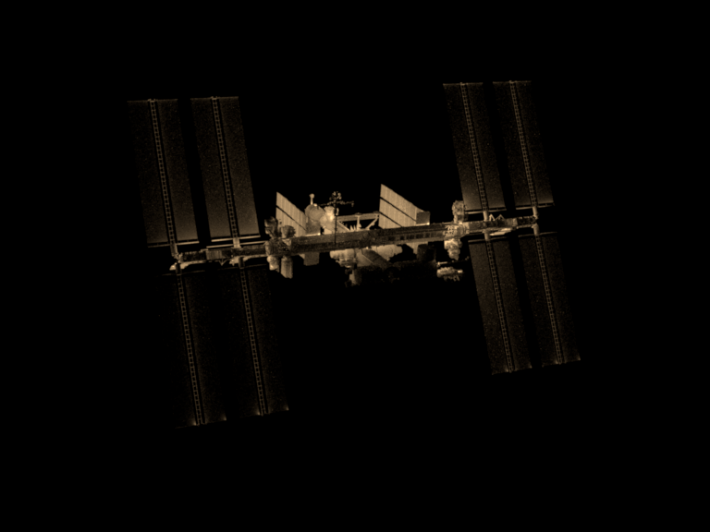}}\qquad
\subfloat[Turbidity level: 6]{\label{fig:ec}\includegraphics[width=0.21\textwidth]{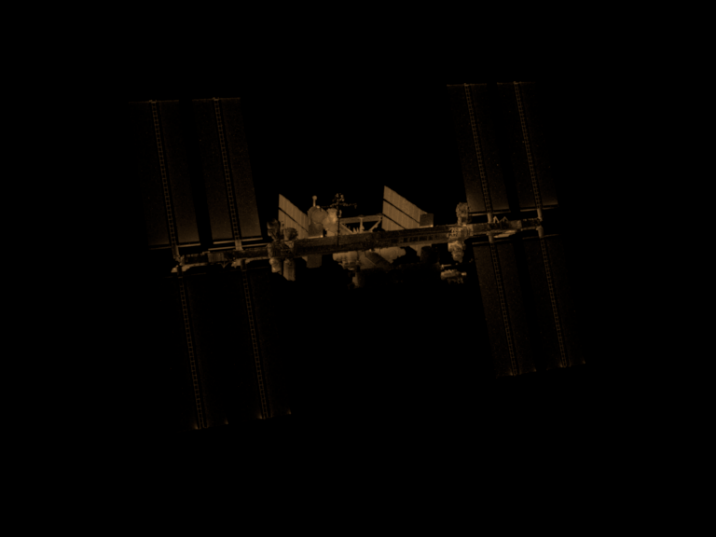}}\qquad%
\subfloat[Turbidity level: 8]{\label{fig:ed}\includegraphics[width=0.21\textwidth]{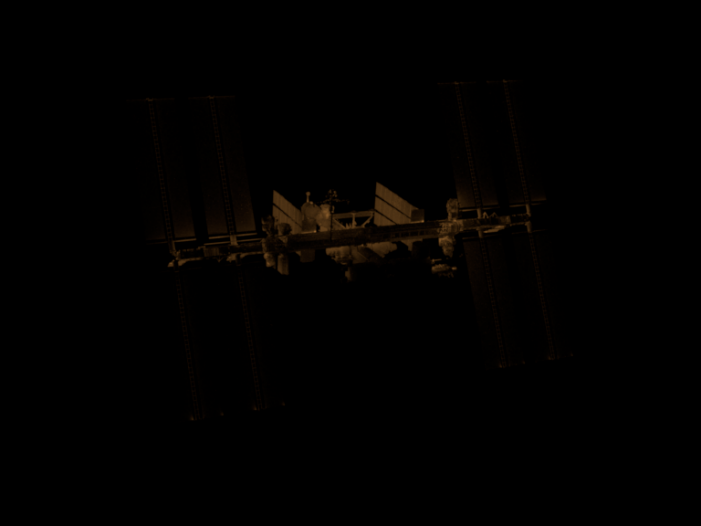}}%
\caption{Ground-based observation of the ISS, illustrating renderings at different values of turbidity.}
\label{fig:exposureSettings}
\end{figure*}

Simulations for ground-based space object observations in the presence of atmospheric scattering effects are delineated through the following rendering exercise. The renderer is enabled to simulate observations at different times of the day, borrowed from the sky-dome radiance model by~\cite{hosek2012analytic}. The daylight renderings of the ISS at different times of the day are shown in Fig.~\ref{fig:timesOfDay}. This exercise emulates the fully focused telescopic observation of the ISS placed in low-earth orbit. The model evaluates the turbidity of the atmosphere and the position of the emitter. {Turbidity} characterizes the scattering of light by the aerosol content in the atmosphere. Increasing the levels of turbidity increases the scattering of light rays and decreases the intensity of radiation. The renderings of the ground-based ISS observations at different levels of turbidity are shown in Fig.~\ref{fig:turbiditySim}. The effects of lower visibility can be offset by increasing the exposure of the virtual camera. The simulation results with variation in the exposure levels are shown in Fig.~\ref{fig:exposureSettings}. In this exercise, the exposure levels are specified in f-stops which, in turn, specify the aperture radius. Higher exposure levels upscale the radiance values applied to the image to result in a brighter output. 

Furthermore, a few other 2D imaging capabilities for aerospace navigation applications are previously demonstrated in~\cite{verras2021vision,bhaskara2023differentiable}.  
Next section focuses on a four-dimensional visual sensing capabilities of the NaRPA through a velocimeter LiDAR model.  

\begin{figure*}[ht!]
\centering
\subfloat[Exposure: $2^0$]{\label{fig:tta}\includegraphics[width=0.21\textwidth]{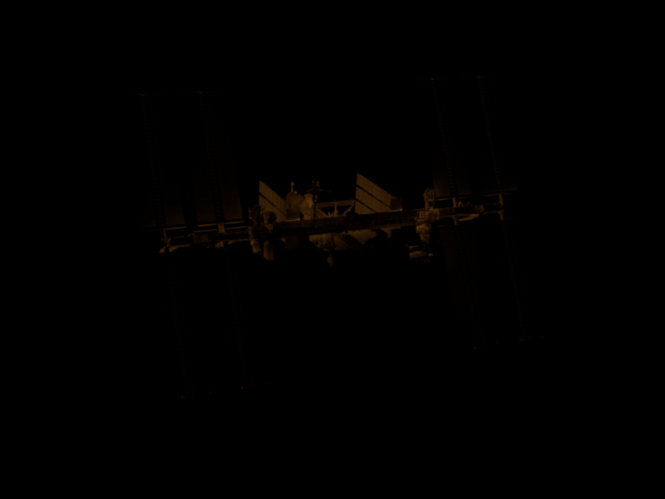}}\qquad
\subfloat[Exposure: $2^1$]{\label{fig:ttb}\includegraphics[width=0.21\textwidth]{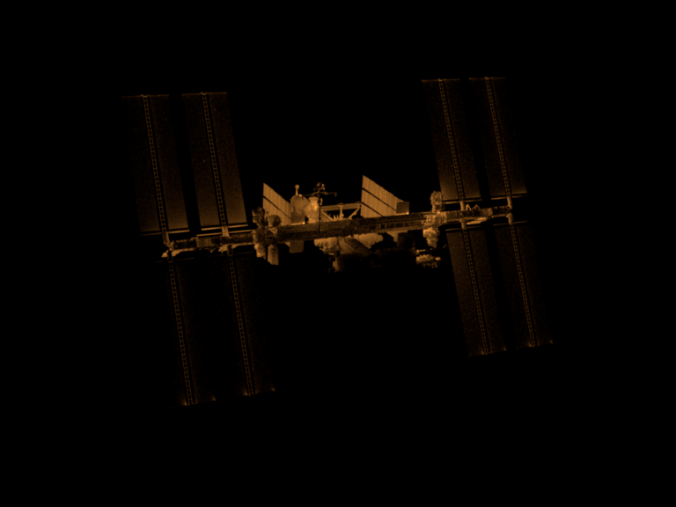}}\qquad
\subfloat[Exposure: $2^2$]{\label{fig:ttc}\includegraphics[width=0.21\textwidth]{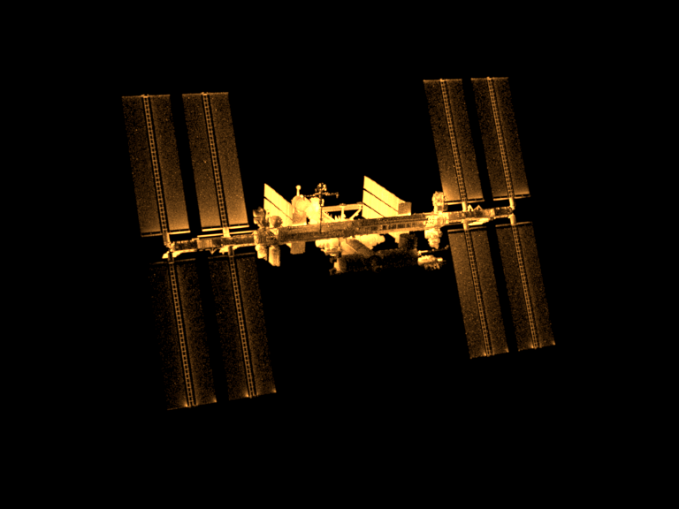}}\qquad%
\subfloat[Exposure: $2^3$]{\label{fig:ttd}\includegraphics[width=0.21\textwidth]{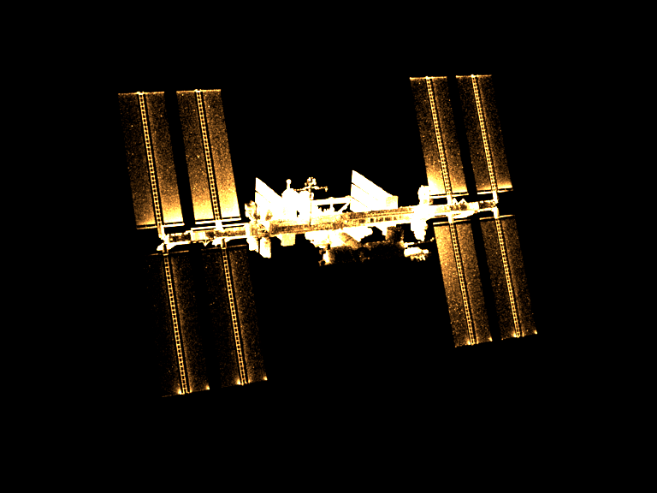}}%
\caption{Ground-based observation of the ISS, illustrating renderings at different values of exposure.}
\comment{
\caption{Ground-based observation of the ISS, illustrating renderings at different values of exposure.}
}
\label{fig:turbiditySim}
\end{figure*}

\comment{
\subsubsection{True data image matching}
An attempt to recreate a real-time telescopic observation of the ISS image from the rendered simulations is shown in Fig. \ref{fig:trueISS}. To match the digital image of the telescopic observation, the exposure, turbidity, and aperture values are iteratively adjusted to render an image that closely resembles the true observation. The scope of this exercise is to identify the parameter values from an observation or a set of time-series observations and to identify the true pose of the ISS from the corrected and improved simulation. In the next section, the pose estimation from a time series of simulated images is described in detail.
 \begin{figure}[ht]
    \centering
   \caption{Pose and parameter estimation from telescopic image.}
    \label{fig:trueISS}
\end{figure}
\FloatBarrier
}

\section{Depth Estimation using Stereo Vision}  \label{sec:stereo}

The computation of depth from 2D images remains one of the topics widely studied in vision-based navigation \cite{scharstein2002taxonomy}. Classical approaches for depth estimation rely on stereo matching via pixel correspondence across images captured from a stereo camera setup. Depths are estimated by triangulation, wherein each corresponding pair of image coordinates is used to estimate the respective 3D coordinates in the scene. Modern approaches utilize deep learning frameworks for depth retrieval and 3D reconstruction tasks \cite{laga2020survey}. In either approach, the availability of high-fidelity ground truth data is monumental in testing and validating the depth estimation algorithms. 

In this paper, {NaRPA is applied as an engine for generating custom resolution stereo imagery}. The rendering engine is capable of generating stereo images for 3D coordinate estimation using calibrated and uncalibrated stereo vision. We demonstrate depth computation in a calibrated stereo application using the classical triangulation approach. As a motivating example, consider the motion of a spacecraft in relative navigation with respect to a planetary terrain, as shown in Fig. \ref{fig: Depth_1}. Two keyframes are observed in the stereoscopic setup, emulating a motion from the stereo application. The coordinate transformation between the two capturing events is assumed to be available from onboard sensors in a terrain-fixed reference frame. 


\begin{figure}[ht!]
    \centering
    \includegraphics[width=0.45\textwidth]{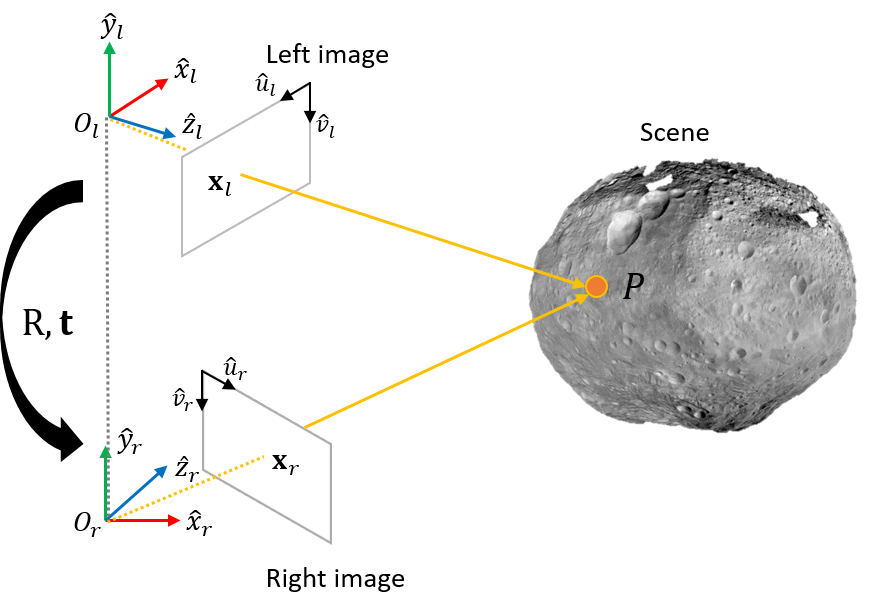}
    \caption{Triangulation to show the relationship between the scene point $P$ and its projections in the left and right camera frames, centered at $O_l$ and, $O_r$ respectively.}
    \label{fig: Depth_1}
\end{figure}

\subsection{Stereoscopic System Model} \label{sec:stereopsis}
Epipolar geometry relates the multiview images of the observed scene to the 3D object in the scene\comment{ \cite{tovsic2009spherical}}. It represents a geometric relation between the spatial coordinates in the scene and their image projections in a stereoscopic setup. Consider an object point $P \in \mathbb{R}^3$ and its two images in left and right camera frames, given by their coordinates $\mathbf{x}_l = [x_l, y_l, z_l]$ and $\mathbf{x}_r = [x_r, y_r, z_r]$. The $\mathbf{x}_l$, $\mathbf{x}_r$, and $P$ are related by the triangular geometry shown in Fig. \ref{fig: Depth_1}. Perspective projection geometry governs the mapping from 3D coordinates onto 2D image plane coordinates, as shown below.

For calibrated pinhole camera sensors with known intrinsics, the projections of $P$ in the left and right image planes are given by
\begin{align} 
    \frac{U_i}{f_x^{(i)}} =\frac{u_i - o_x^{(i)}}{f_x^{(i)}} = \frac{x_i}{z_i} \label{eq:dest_U_pix_eq}
    \\
   \frac{V_i}{f_y^{(i)}} = \frac{v_i - o_y^{(i)}}{f_y^{(i)}} = \frac{y_i}{z_i} \label{eq:dest_V_pix_eq}
\end{align}

\noindent where $i$ is a placeholder to indicate the coordinate frames of the left and right cameras ($l$ and $r$). $u_i$ and $v_i$ are the image projections of $P$ in the left and right image planes. $U_i$ and $V_i$ are the shorthand notation for image-centered pixel coordinates. Focal lengths ($f_x^{(i)}, f_y^{(i)}$) and principal point offsets ($o_x^{(i)}, o_y^{(i)}$) denote the camera intrinsic parameters of the left and right cameras.

The homogeneous coordinates $\mathbf{x}_l$ and $\mathbf{x}_r$ are related by a known transformation between the left and right camera frames as 
\begin{equation} \label{eq:dest_coord_transform}
    \begin{bmatrix}
    x_r \\ y_r \\ z_r \\ 1
    \end{bmatrix} =
    \begin{bmatrix}
    R_{11} & R_{12} & R_{13} & t_x
    \\
    R_{21} & R_{22} & R_{23} & t_y \\
    R_{31} & R_{32} & R_{33} & t_z \\
    0 & 0 & 0 & 1
    \end{bmatrix}
    \begin{bmatrix}
    x_l \\ y_l \\ z_l \\ 1
    \end{bmatrix}
\end{equation}

The division of $x_r$ with $z_r$ in Eq. (\ref{eq:dest_coord_transform}) yields the expression on the right-hand side of the perspective projection equation (\ref{eq:dest_U_pix_eq}) 
\begin{align}
    \frac{x_r}{z_r} = \frac{R_{11} x_l  + R_{12}y_l  + R_{13}z_l + t_x}{R_{31} x_l  + R_{32}y_l + R_{33} z_l  + t_z} 
    \\
    \frac{x_r}{z_r} =  \frac{R_{11} \frac{x_l}{z_l}   + R_{12}\frac{y_l}{z_l}  + R_{13} + \frac{t_x}{z_l}}{R_{31}\frac{x_l}{z_l}  + R_{32} \frac{y_l}{z_l}  + R_{33} + \frac{t_z}{z_l}} 
\end{align}
Now, substituting the perspective projection equations (\ref{eq:dest_U_pix_eq} \& \ref{eq:dest_V_pix_eq}), we obtain the following. 
\begin{align}
    \frac{U_r}{f_x^{(r)}} =  \frac{R_{11} \frac{U_l}{f_x^{(l)}}  + R_{12} \frac{V_l}{f_y^{(l)}} + R_{13} + \frac{t_x}{z_l}}{R_{31} \frac{U_l}{f_x^{(l)}}  + R_{32} \frac{V_l}{f_y^{(l)}} + R_{33} + \frac{t_z}{z_l}} 
\end{align}

Upon simplification, $z_l$, the $z$ coordinate of $P$ in the left camera frame, is expressed as
\begin{equation}
    z_l = \frac{t_z U_r - f_x^{(r)} t_x}{ \left(R_{11} f_x^{(r)} - R_{31}U_r \right) \frac{U_l}{f_x^{(l)}} + \left(R_{12} f_x^{(r)} - R_{32}U_r\right) \frac{V_l}{f_y^{(l)}}   + \left(R_{13} f_x^{(r)}-R_{33}U_r \right) }
    \label{eq:dest_zl}
\end{equation}

Using the projection equations again, the $x$ and $y$ coordinates of $P$ in the left camera frame are recovered as 
\begin{align}
    x_l = z_l \frac{U_l}{f_x^{(l)}} \label{eq:dest_xl}
    \\
    y_l = z_l \frac{V_l}{f_y^{(l)}} \label{eq:dest_yl}
\end{align}

\comment{

For calibrated pinhole camera sensors with known intrinsics ($K_l$ and $K_r$), the projections of $P$ in the left and right cameras are given by
\begin{align}
    \mathbf{p}_l &= K_l \mathbf{x}_l \label{eq:depthEst_p1}
    \\
    \mathbf{p}_r &= K_r \mathbf{x}_r \label{eq:depthEst_p2}
\end{align}

\noindent where $\mathbf{p}_i = [u_i \quad v_i \quad 1]$ ($i = l, r$) are the image projections of $P$ in pixels, $K_i$ is a camera intrinsic matrix of focal lengths ($f_x, f_y$), principal point offset ($o_x, o_y$) given by $K_i = \begin{bmatrix}
f_x^{(i)} & 0 & o_x^{(i)} \\ 0 & f_y^{(i)} & o_y^{(i)} \\ 0 & 0 & 1 \end{bmatrix}$

The homogeneous coordinates $\mathbf{x}_l$ and $\mathbf{x}_r$ are related by a known transformation between the left and right image frames, as 
\begin{equation}
    \mathbf{x}_r = \begin{bmatrix}
    R & \mathbf{t} \\ \mathbf{0}_{1 \times 3} & 1
    \end{bmatrix} \mathbf{x}_l
\end{equation}
\noindent substituting for $\mathbf{x}_r$, Eq. (\ref{eq:depthEst_p2}) is rewritten as
\begin{equation} \label{eq:depEst_p12}
     \mathbf{p}_r = \underbrace{K_r \begin{bmatrix}
    R & \mathbf{t} \\ \mathbf{0}_{1 \times 3} & 1
    \end{bmatrix}}_{M_r} \mathbf{x}_l = M_r \mathbf{x}_l 
\end{equation}

Using the geometric relationship between the left and the right camera frames along with their corresponding pixel projections, the 3D scene coordinates $\mathbf{x}_l$ , in the left camera frame (or $\mathbf{x}_r$, if parameterized, in the right camera frame) are determined. 

The geometric relationships between $\mathbf{x}_l$ and the corresponding pixel projections, shown in Eqs. (\ref{eq:depthEst_p1}) and (\ref{eq:depEst_p12}), are utilized to obtain a linear system of four equations in three unknowns. Using linear least squares, the 3D scene coordinates $\mathbf{x}_l$, in the left camera frame (or $\mathbf{x}_r$, if parameterized in the right camera frame), are estimated.
}
Using the geometric relationship between the left and the right camera frames along with their corresponding image plane projections, the 3D scene coordinates $\mathbf{x}_l$ , in the left camera frame (or $\mathbf{x}_r$, if parameterized, in the right camera frame) are determined.

\subsection{Results}
\subsubsection{Scene Configuration}

In this exercise, two imaging events in a planetary descent simulation are rendered for 3D coordinate estimation using triangulation. Two monocular images of a terrain are captured at two deterministic poses in the virtual scene geometry, with the terrain under sufficient illumination. The geometric parameters for the image are captured in Table \ref{tab:cameraSetup}. Optical specifications include pinhole camera resolution of $500 \times 500$ pixels, diagonal field of view of $30^\circ$. Computation of 3D feature coordinates is performed according to the following principles: 

\begin{itemize}
    \item The image correspondence through stereo matching is established through sparse corner features extracted using Eigenvalue-based geometric feature detection \cite{{weinmann2014semantic}}. The matching stereo pairs $\mathbf{p}_l$ and $\mathbf{p}_r$ are acquired from the descriptors of the corner features detected in the key frames.  
    \item 3D terrain coordinates are realized by computing the expressions in Eqs. (\ref{eq:dest_zl}, \ref{eq:dest_xl}, \ref{eq:dest_yl}).
    \item For verification, ground truth is established from the generated point cloud of the object (see Section \ref{sec:rayTracerCap}) as well as the known transformation from individual camera frames to the terrain-fixed reference frame. 
\end{itemize}

\begin{table}[ht]
\centering
\begin{tabular}{ |c|c|c| } 
 \hline
Target (m) & Origin (m) & Up (rad) \\ 
\hline
 $[0 \quad 0 \quad 0]^T$ & $[0 \quad 0 \quad 800]^T$ & $[1 \quad 0 \quad 0]^T$ \\ 
 $[0 \quad -30 \quad 0]^T$ & $[0 \quad -30 \quad 750]^T$ & $[1 \quad 0 \quad 0]^T$ \\ 
 \hline
\end{tabular}
\caption{ \label{tab:cameraSetup} Geometric parameters for camera positioning and orientation in the scene: `Target' indicates the camera pointing location, `Origin' indicates the camera position, and `Up' represents the camera alignment.}
\end{table}

\subsubsection{Results}
3D coordinates are estimated in the left camera frame ($\hat{\mathbf{x}}_l$) and the percentage of relative errors along each of the coordinates is calculated as 
\begin{equation}
    \% \text{Error} = \frac{|\hat{\mathbf{x}}_{l,i} - \mathbf{x}_{l,i}|}{|\mathbf{x}_{l,i}|} \times 100; \hspace{1cm} i=1,2,3
\end{equation}

The estimated 3D coordinates are re-projected into the left camera frame using the known camera intrinsics. The estimated and true pixel coordinates are highlighted in Fig. \ref{fig:dest_worst20}. Figure \ref{fig:dest_absRelErrors} shows the distribution of absolute errors and the percentage of relative errors in the estimation of 3D coordinates. The maximum absolute errors are observed to be along the $z$ coordinates, which is expected due to the placement of the camera relatively further along $z$. Outliers in the percentage of relative errors are more along $y$ because the $y$ coordinates of these feature points are very close to $0$ (with a significant quotient in the division of two values of close floating points). 

\begin{figure*}[t]
  \subfloat[True and estimated pixel projections (bottom 20 matches)]{\includegraphics[width=0.45\textwidth]{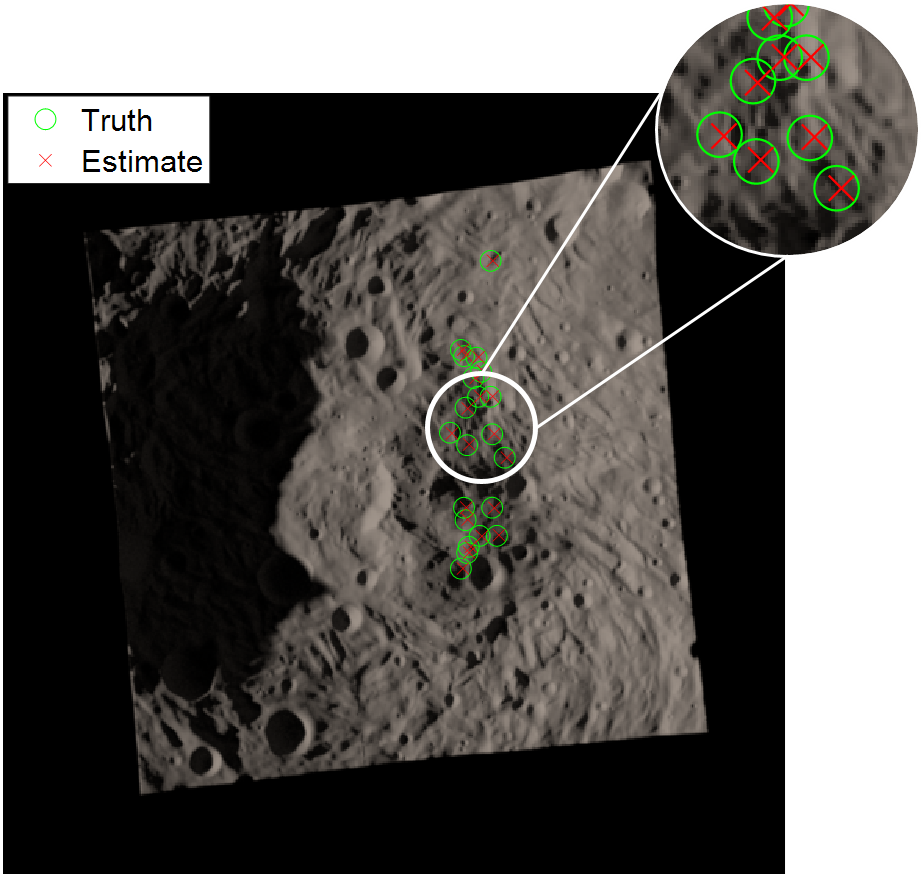}{\label{fig:dest_worst20}}}
  \hfill
  \subfloat[Errors in the $x,y,z$ estimates evaluated for 100 feature points. Outliers are indicated by \textcolor{red}{+}.]{\includegraphics[width=0.45\textwidth]{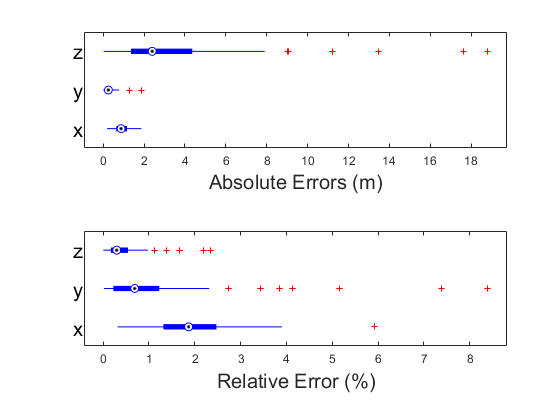}{\label{fig:dest_absRelErrors}}}
  \caption{Results of 3D coordinate estimation: On the left, the estimated coordinates are reprojected into the camera frame for visual comparison. On the right, absolute, and relative error distributions between the estimated and the true coordinates are portrayed.}
\end{figure*}


In this capability demonstration, the 3D coordinates are not estimated to subpixel accuracy. However, using dense point correspondences from renderings, effective outlier rejection schemes, and efficient pixel interpolation schemes, applications such as structure from motion and 3D scene reconstruction are realizable.

\section{Velocimeter LiDAR Simulation} \label{sec:velocimeterLidar}
A velocimeter LiDAR\footref{fn:velocimeterLIDAR} provides direct observation of relative velocity and is an important navigation instrument in precision landing~\citep{wolf2006performance,adams2022velocimeter}. 
This LiDAR obtains precise range and velocity information by capturing the Doppler frequency shift between an emitted and returned frequency-modulated continuous wave~\citep{amzajerdian2012doppler}. 
The simulated sensor model is a scanning type LiDAR that generates range vectors to each target in the sensor's field of view. 
Using the knowledge of relative velocity between a target and the sensor, line-of-sight (LOS) velocity measurements for each target in the field of view and thereby emulate an FMCW Doppler LiDAR.

This section describes a functional model of ray-tracing-based velocimeter LiDAR.\footnote{Portions of the velocimeter LiDAR model and the experimental validation reported in this section were previously presented at the AIAA SciTech 2025 Forum~\citep{bhaskara2025development}.\label{fn:priorpub}}
The virtual LiDAR is capable of synthesizing a stream of point cloud measurements and the velocimetry data in a terrain-relative navigation pipeline. By taking real-world measurements obtained using a velocimeter LiDAR system as a reference, synthetic measurements obtained using the sensor model are compared. This enables calibration of the sensor model to match up to an on-field sensor and to evaluate the fidelity of virtual sensing modalities in a simulation environment.

\subsection{Description of Doppler Velocity Model}

\begin{figure}[ht]
    \centering
    \includegraphics[width=.5\textwidth]{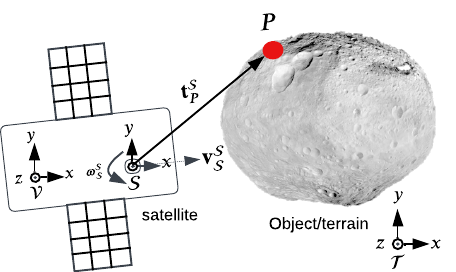}
    \caption{Illustration of coordinate frames, translation, and velocity vectors. LiDAR sensor (frame $\mathcal{S}$) is mounted on the satellite vehicle (frame $\mathcal{V}$) to scans objects in its field of view. $P$ is a static point on the object (frame $\mathcal{T}$). \scitechfig}
    \label{fig:Lidar_coordinateSys}
\end{figure}

The coordinate frames in this description are illustrated in Fig.~\ref{fig:Lidar_coordinateSys}. 
Assuming that the LiDAR sensor is mounted on a rigid vehicle and that the transformation from the sensor frame $\{\mathcal{S}\}$ to the vehicle frame $\{\mathcal{V}\}$ is known beforehand from extrinsic calibration through a translation vector $\mathbf{t}^{\mathcal{V}}_{\mathcal{S}}$ and a rotation matrix $\mathbf{R}^{\mathcal{V}}_{\mathcal{S}}$.  

If the vehicle's translational and angular velocities are identified in the vehicle's frame to be $\mathbf{v}_{{V}}^{\mathcal{V}}$ and $\boldsymbol{\omega}_{{V}}^{\mathcal{V}}$, then the velocity of the LiDAR sensor $\mathbf{v}_{\mathcal{S}}^{\mathcal{V}}$ and its angular velocity $\boldsymbol{\omega}_{\mathcal{S}}^{\mathcal{V}}$ expressed in the vehicle frame are computed using transport theorem as:
\begin{equation}
    \mathbf{v}_{\mathcal{S}}^{\mathcal{V}} = \mathbf{v}_{\mathcal{V}}^{\mathcal{V}} + \boldsymbol{\omega}_{\mathcal{V}}^{\mathcal{V}} \times \mathbf{t}^{\mathcal{V}}_{\mathcal{S}}
\end{equation}

For a point $P$  detected in the LiDAR's field of view with $\mathbf{t}_{P}^{\mathcal{S}}$ being a vector from origin of $\{\mathcal{S}\}$ to $P$, its relative velocity   $\dot{\mathbf{t}}_{P}^{\mathcal{S}}$, representing the rate of change of the range measurement, is expressed as: 
\begin{equation}
    \mathbf{v}_{P}^{\mathcal{S}} = \mathbf{v}_{\mathcal{S}}^{\mathcal{S}} + \dot{\mathbf{t}}_{P}^{\mathcal{S}} +  \boldsymbol{\omega}_{\mathcal{S}}^{\mathcal{S}} \times \mathbf{t}_{P}^{\mathcal{S}}
\end{equation}
\noindent where $\mathbf{t}_{P}^{\mathcal{S}}$ is also the range measurement of the LiDAR and $\{\mathbf{v}_{{S}}^{\mathcal{S}}, \boldsymbol{\omega}_{\mathcal{S}}^{\mathcal{S}}\}$ denotes the velocities of the LiDAR in the LiDAR sensor frame. 
Also, since the $P$ is stationary with respect to the terrain (object) frame ${\mathcal{T}}$, $\mathbf{v}_{P}^{\mathcal{S}} = \mathbf{0}$. Then the relative velocity (range rate) of $P$ relative to the LiDAR frame is:
\begin{equation} \label{eq:relVelocityEq}
    \dot{\mathbf{t}}_{P}^{\mathcal{S}} = - \mathbf{v}_{{S}}^{\mathcal{S}} - \boldsymbol{\omega}_{{S}}^{\mathcal{S}} \times \mathbf{t}_{P}^{\mathcal{S}}
\end{equation}

The Doppler sensors only measure the relative velocities in the radial direction, i.e., along their line-of-sight from the origin, which is along the direction of $\mathbf{t}_{P}^{\mathcal{S}}$.  
Let us denote the direction vector from $\{\mathcal{S}\}$ to $P$ is denoted as $\hat{\mathbf{r}}_P$ such that 
\begin{equation}
    \hat{\mathbf{r}}_P = \frac{\mathbf{t}_{P}^{\mathcal{S}}}{||\mathbf{t}_{P}^{\mathcal{S}}||}
\end{equation}

Then the instantaneous Doppler velocity ($v_P$), representing the radial or line-of-sight velocity of the point $P$ relative to the LiDAR sensor frame, is the projection of the relative velocity vector  $\dot{\mathbf{t}}_{P}^{\mathcal{S}}$ along $\hat{\mathbf{r}}_P$ and using Eq. (\ref{eq:relVelocityEq}) as 
\begin{align}
    v_P &= \hat{\mathbf{r}}_P \cdot \dot{\mathbf{t}}_{P}^{\mathcal{S}}  \nonumber
    \\
    &= - \hat{\mathbf{r}}_P \cdot ( \mathbf{v}_{{S}}^{\mathcal{S}} + \boldsymbol{\omega}_{{S}}^{\mathcal{S}} \times \mathbf{t}_{P}^{\mathcal{S}} )
\end{align}

Based on the sensor kinematics described above, the actual measurements from a LiDAR in a monostatic configuration (where the transmitter and receiver are co-located) is a Doppler shift in frequency that is given by 
\begin{equation}
    \Delta f = \frac{2 v_{P}} {c} f = \frac{2 v_P}{\lambda}
\end{equation}
\noindent where $c$ is the velocity of light, $f$ and $\lambda$ are the frequency and wavelength of the continuously transmitted laser.

In a LiDAR scan pass, NaRPA renders point cloud and Doppler velocities in a sensor frame. 
The sensor kinematics connect the relative motion to the generated data packet at every scan instance.  
The kinematics is particularly important to translate a real world experiment into a virtual scene.
The scene model along with the motion description can be imported into the NaRPA engine for LiDAR scans along the trajectory.
NaRPA returns the point clouds metadata at a set frame rate of the sensor. 
The simulation architecture is described below.

\subsection{Sensor simulation}
A non-recursive ray tracing model operates by casting rays into each pixel in a sensor's view frame and lends itself to the utility of a LiDAR sensor's functional model. Scanning is achieved by successively shooting light rays with a specified angular resolution in the entirety of the LiDAR's field of view. This enables the LiDAR sensor model to map the virtual world through point cloud measurements as output. 

\begin{figure*}[ht]
    \centering
    \includegraphics[width=.75\textwidth]{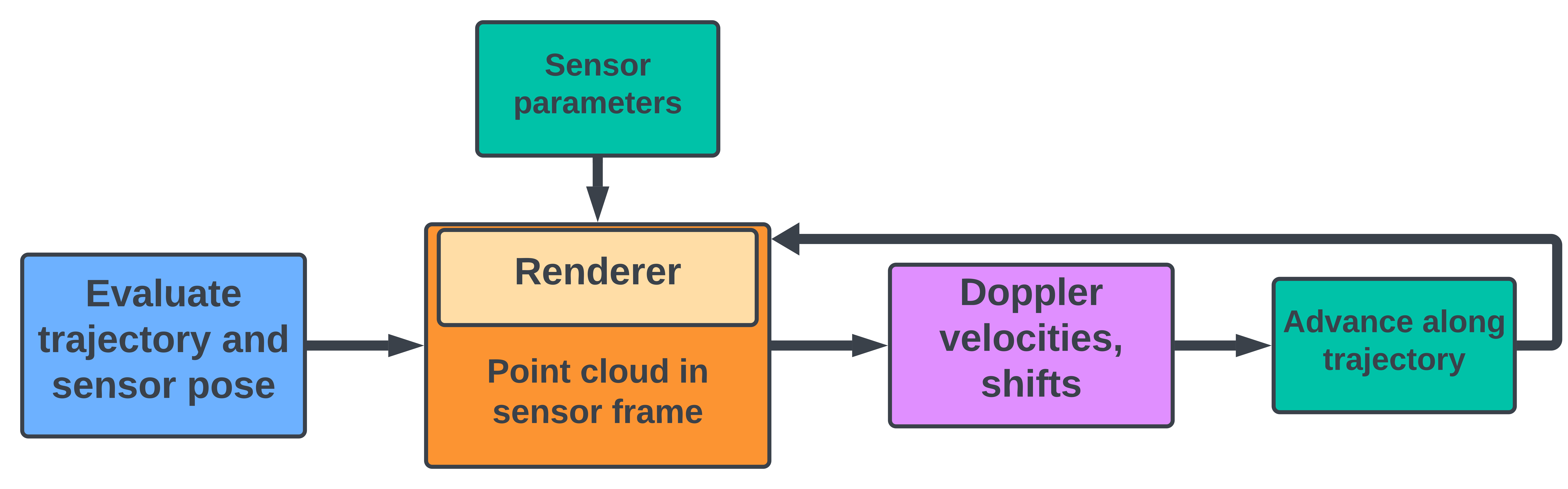}
    \caption{Velocimeter LiDAR simulation pipeline. \scitechfig}
    \label{fig:sim_pipeline}
\end{figure*}

The sensor simulation pipeline is shown in Fig.~\ref{fig:sim_pipeline}. For given sensor optics and trajectory metadata including vehicle position and velocity, the NaRPA engine generates point clouds and frequency shift measurements with LOS velocities in simulation time. With the process tunable parameters such as laser wavelength, frame rate, angular resolution, and noise., the rendering engine is capable of generating point clouds of desired densities. 
The 3D points and the Doppler velocity measurements may be subject to various sources of errors such as sensor noise, calibration errors, or assumptions about the target's motion, which could all be accounted for in the simulations as a stochastic noise process.




\subsection{Experimental Results} \label{sec:s3_experiments}



An accurate representation of real-world objects and corresponding relative motion is important for virtual dataset generation that closely emulate the realism.
In this pipeline, the process of converting real-world data into a virtual scene for accurate simulation involves (a) 3D asset generation and (b) ground truth sensor trajectory. 
Firstly, if the 3D model of the real-object is unavailable, an RGB camera colocated with a LiDAR captures the dynamic scene. 
The camera-LiDAR coordinate transformation is calibrated beforehand.
The camera's 2D imagery is then used for dense 3D reconstruction using photogrammetry software. COLMAP~\citep{schoenberger2016vote} is used in this pipeline. 
The reconstructed 3D object is converted into textured meshes in compatible OBJ format using software such as MeshLab~\citep{Cignoni2008MeshLabAO}. 
Secondly, to recreate the sensor's motion relative to the 3D asset, ground truth trajectory information is obtained from motion capture (MoCap) systems or coarsely estimated from COLMAP's camera pose data. 
MoCap provides accurate velocity data which can also be coarsely estimated from  COLMAP if the frame rate is sufficiently high. 
Using the camera intrinsics, the angular and linear velocities ($\boldsymbol{\omega}_V^{\mathcal{V}}$ and $\mathbf{v}_V^{\mathcal{V}}$) within one LiDAR sample time $\Delta t$ can be approximately computed as 
\begin{align}
    \boldsymbol{\omega}_V^{\mathcal{V}} &= \frac{\mathbf{u}_\theta}{\Delta t}
    \\
    \mathbf{v}_V^{\mathcal{V}} &= \frac{\mathbf{u}_t}{\Delta t}
\end{align}

In essence, 2D imagery enables reconstruction of 3D scene using structure-from-motion algorithms while trajectory information is extracted from motion capture metadata.
This process enables the creation of virtual datasets that closely emulate real-world scenarios, crucial for emulating the real-world scene and the sensor motion using the NaRPA engine.

For hardware experiments, Texas A\&M's Land, Air, and Space robotics (LASR) lab spinning satellite testbed is used to acquire the velocimetry and point cloud data in a space emulating environment. 
The testbed features a fully equipped Vicon\footnote{Vicon motion capture system: \url{https://www.vicon.com/}, accessed on September 9, 2024.} motion capture system to obtain the ground truth relative motion kinematics. 
The spinning satellite testbed is a dynamic scene as indicated in Fig.~\ref{fig:spinningRocket}. 
The satellite module rotates at a pre-defined angular velocity for the LiDAR and camera sensors to capture the scene. 
The Doppler velocity measurement are captured from a static planar robot with the satellite module in the LiDAR's field of view.





To validate the LiDAR model implemented on NaRPA engine, comparisons are made between the experimental data and rendered simulation data. 
For statistical analysis, the spinning satellite dataset is used to render point clouds and point velocities during an observation sequence in LiDAR's reference frame.
Point clouds and velocities from the commercial Aeva\footnote{Aeva is a commerical FMCW LiDAR manufacturer: \url{https://www.aeva.com/}, accessed on September 11, 2024.\label{fn:aeva}} LiDAR are plotted in Fig.~\ref{fig:aeva_rocket} while the simulated measurements are shown in Fig.~\ref{fig:narpa_rocket}.
A qualitative match between the point ranges and Doppler velocities from the respective real and the virtual scenes is visually indicated. 
Note that the number of the points in this comparison are different. Reasons being (1) the real-world measurements may not return all the visible points due to surface properties, (2) dense reconstruction yields meshes with interpolated surfaces, and (3) NaRPA's shortcoming in simulating vertical and horizontal scan resolution accurately.

\begin{figure*}[ht]
     \centering
     \subfloat[][Frame from spinning satellite experiment.]
     {\includegraphics[width=.318\textwidth]{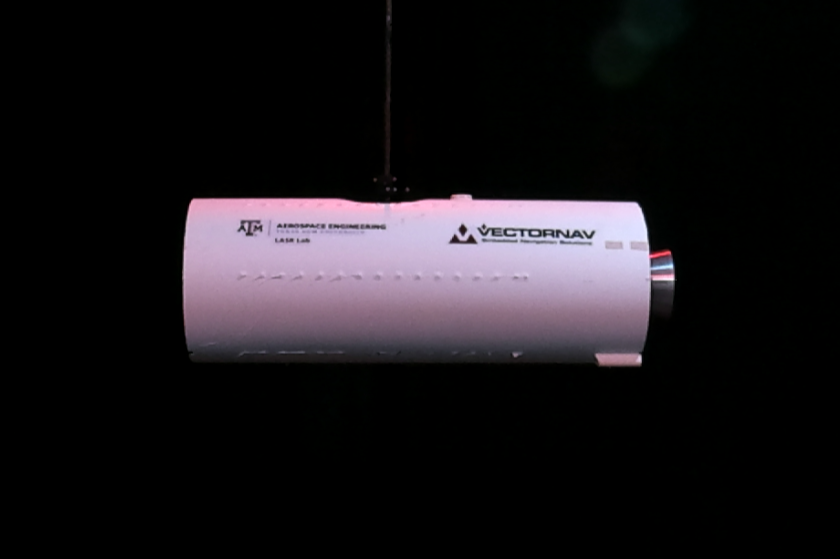}\label{fig:spinningRocket}} 
     \hfill
     \subfloat[][Point cloud and point velocities from commercial (Aeva\footref{fn:aeva}) Doppler LiDAR.]{\includegraphics[width=.33\textwidth]{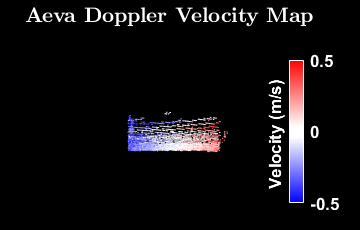}\label{fig:aeva_rocket}}
     \hfill
     \subfloat[][Reconstructed Point cloud and velocities from simulation.]{\includegraphics[width=.33\textwidth]{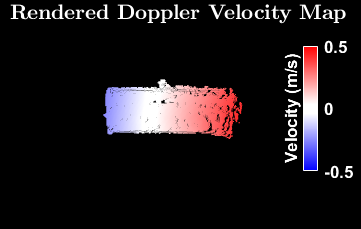}\label{fig:narpa_rocket}} 
     \caption{Comparison of point clouds and instantaneous velocities between real measurements and simulation rendered at same frame. \scitechfig}
     \label{fig:validation_lidar}
\end{figure*}

Because the number of points in the real and the simulated scenes differ, the bulk statistics of the measurements are analyzed for statistical correctness of the simulations. To achieve this, the point velocity distributions across various frames are compared using normalized histograms. 
Fig.~\ref{fig:frame24} presents the normalized histogram of point velocities for a single frame, illustrating the relative frequency of velocities within specified bins. This provides an immediate visual comparison of the velocity distribution between the real and simulated data for that frame.

To further validate the statistical accuracy, the analysis is extended to include the averaged velocity distribution over 100 frames, as shown in Fig.~\ref{fig:average_100frames}. This averaged distribution smooths out frame-to-frame variations, offering a more comprehensive view of the overall velocity distribution. By comparing these normalized histograms, the fidelity of the simulated data against real-world measurements is assessed, ensuring that the simulations accurately represent the dynamic behavior of the scene.

\begin{figure}[ht!]
     \centering
     \subfloat[][Doppler velocities and corresponding rendered point cloud velocities for a chosen single frame.]
     {\includegraphics[width=.48\textwidth]{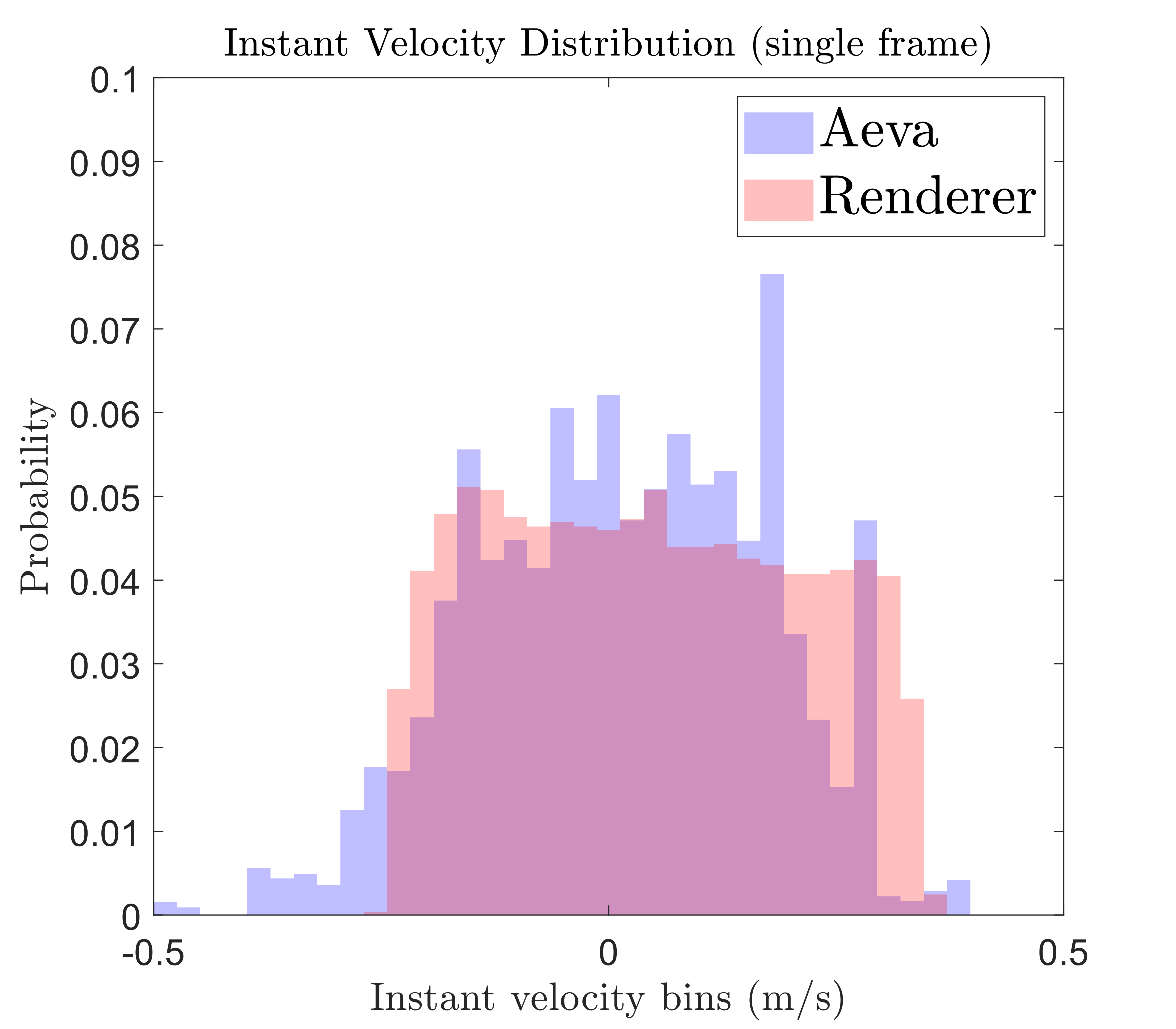} \label{fig:frame24}}
     \\
     \subfloat[][Doppler velocities and corresponding rendered point cloud velocities averaged across 100 frames.]
     {\includegraphics[width=.48\textwidth]{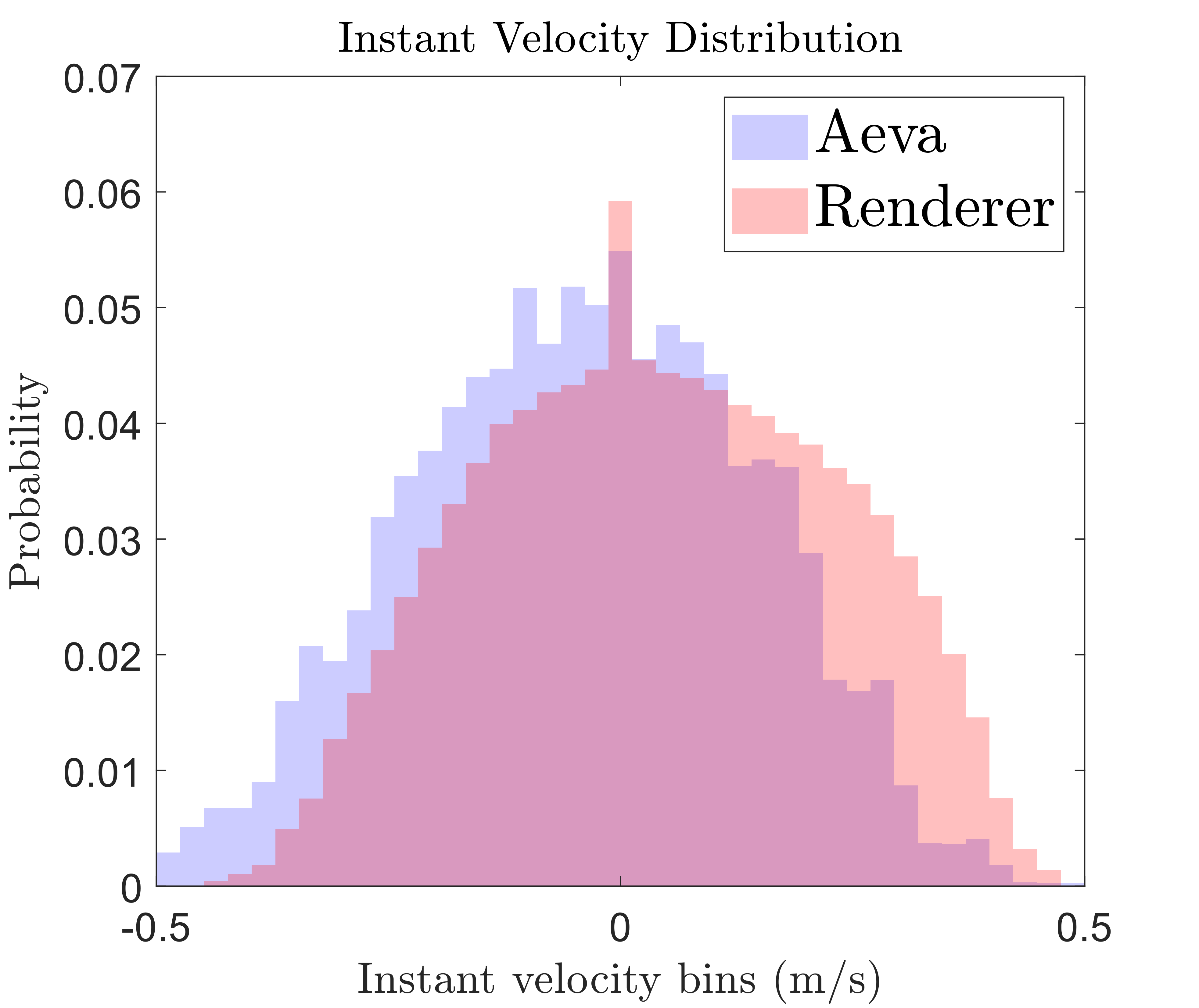}\label{fig:average_100frames}}
     \caption{Normalized histograms of Doppler velocities as a metric to quantify
the distribution of instantaneous point velocities. \scitechfig}
     \label{fig:histograms}
\end{figure}

In addition, median absolute deviation (MAD) of point velocities is chosen as a metric to quantify the instantaneous velocity deviations across frames in a single run. Median absolute deviation indicates the variability of the point velocities, providing a more resilient measure of dispersion compared to standard deviation, especially in the presence of outliers. This metric helps in understanding the consistency and reliability of the velocity estimates.
From Fig.~\ref{fig:MAD}, it is evident that the rendered point velocities closely follow the measured data within the $1-\sigma$ bounds of the simulated measurement. This alignment within the $1-\sigma$ bounds demonstrates that the rendered velocities are statistically consistent with the real measurements.

Furthermore,  the variance of the simulated point cloud, represented in Fig.~\ref{fig:narpa_rocket}, is compared with the points returned from the LiDAR. The un-normalized variances of the point clouds for the rendered points and the Aeva points are shown in Fig.~\ref{fig:sampleVar}. The matching shapes of the variances indicate a high degree of similarity in the spreadedness of the point clouds across frames. This similarity demonstrates that the simulated point cloud closely replicates the distribution and dispersion characteristics of the real LiDAR data, underscoring the fidelity of the proposed NaRPA engine.
However, the open-loop validation studies also indicate that significant improvements can be made to the sensor model in accurately capturing scan resolution and noise properties. 
Additional tuning of the sensor model and closed-loop validation studies are intended for future work.

\begin{figure}[ht!]
    \centering
     \includegraphics[width=.48\textwidth]{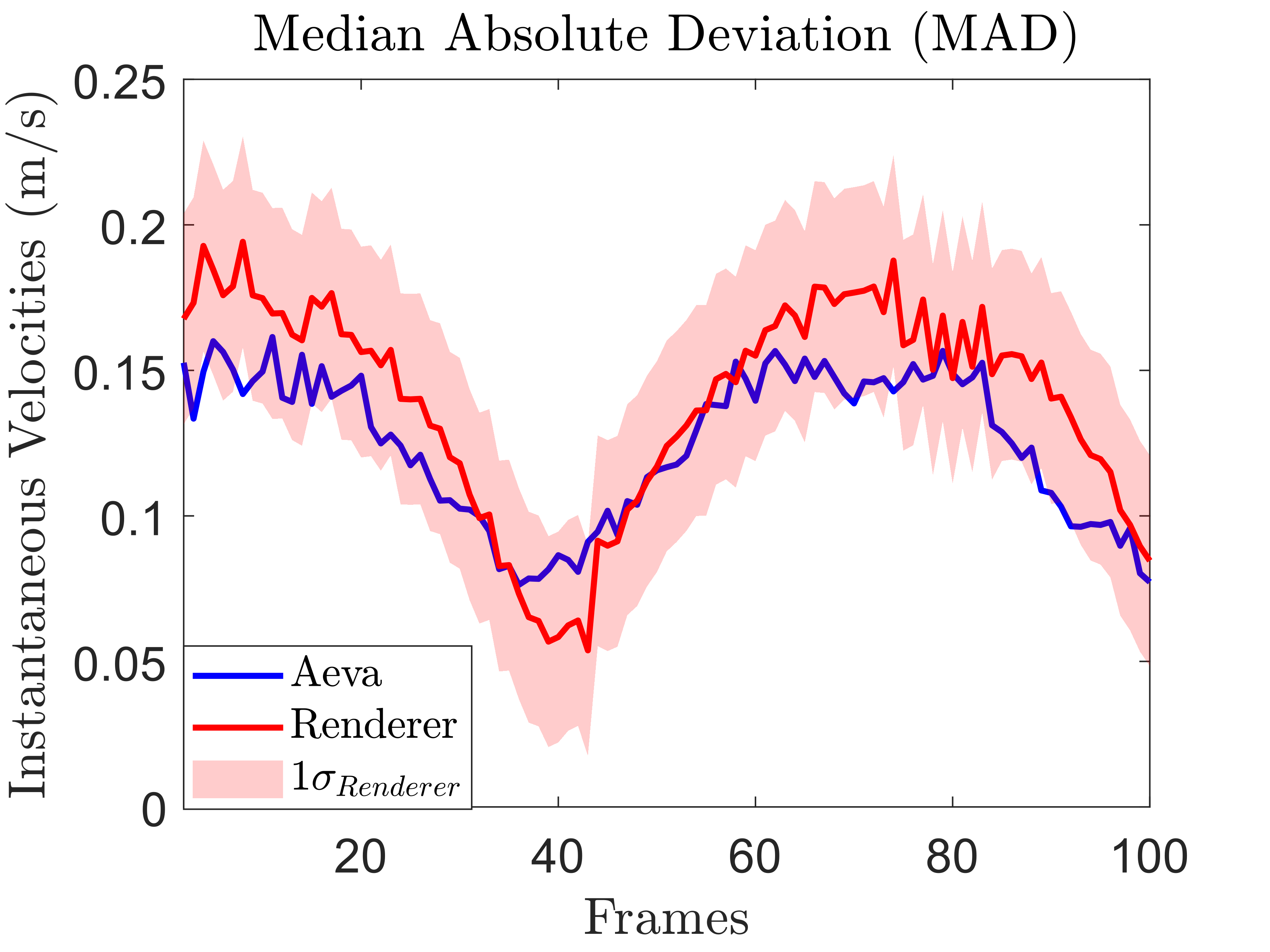}
    \caption{Median Absolute Deviations (MAD) of point velocities from NaRPA (red) and Aeva (blue). Renderer following the velocity distribution of the sensor within 1-sigma bounds. \scitechfig}
    \label{fig:MAD}
\end{figure}

\begin{figure}[ht!]
    \centering
    \includegraphics[width=.48\textwidth]{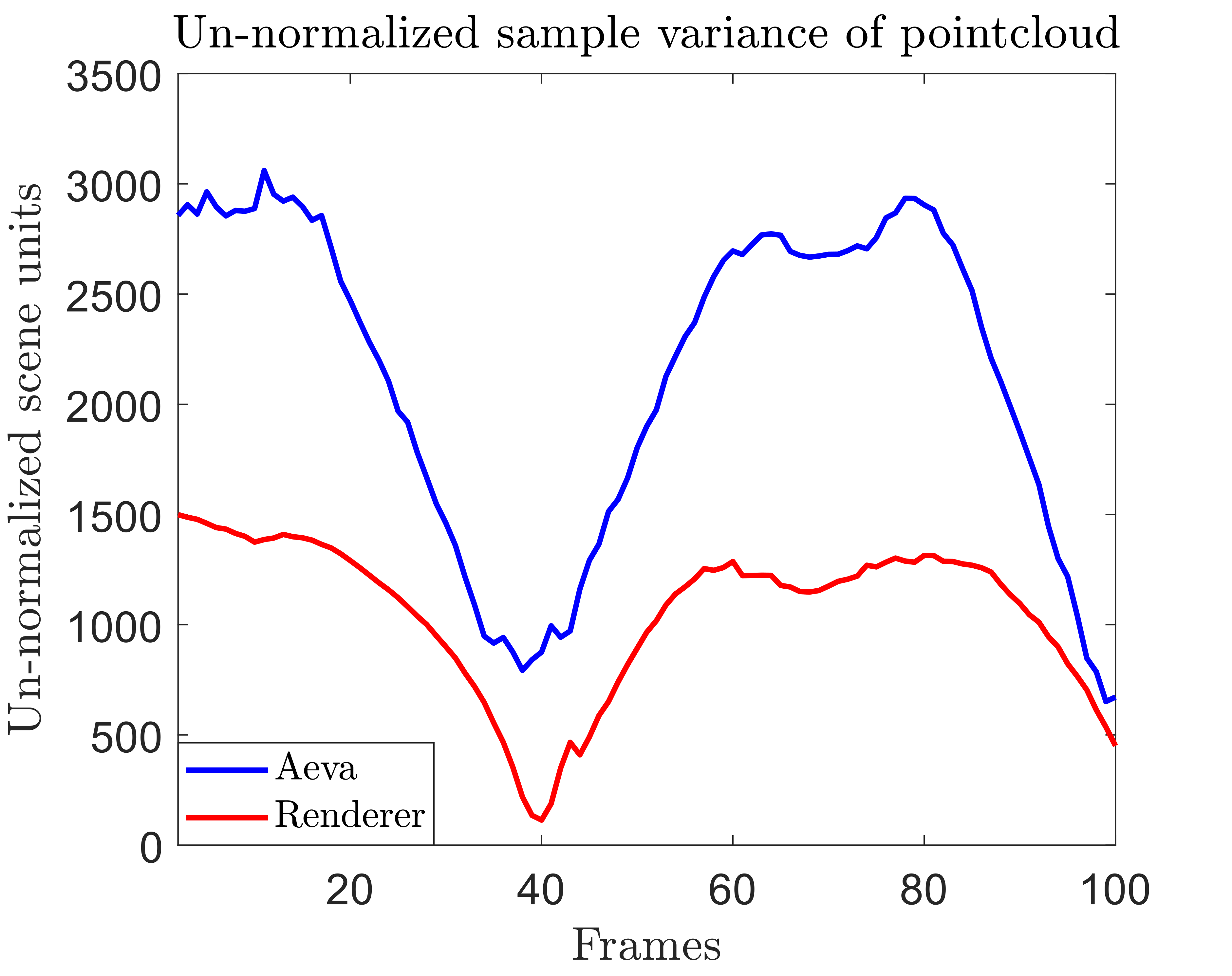}
    \caption{Un-normalized variances of point clouds of rendered points (red) and Aeva points (blue). Matching shapes indicate similarity in point cloud “spreadedness” across frames. \scitechfig}
    \label{fig:sampleVar}
\end{figure}

\section{Discussion} \label{sec:s5_discussion}

NaRPA is a ray-tracing engine designed to produce physically-accurate sensor simulations for space-borne imaging. 
Initially developed for simulating imagery and velocimeter LiDAR for Entry, Descent, and Landing (EDL) operations, NaRPA is evolving into a comprehensive tool for spacecraft visual navigation simulation. 
However, several limitations need to be addressed to enhance NaRPA's capabilities as a high-fidelity tool for multi-mission modeling. 
First, while it strongly emphasizes scientific rendering, NaRPA does not natively provide custom BRDFs for regolith-covered planetary surfaces. 
However, custom BRDF models can be developed and integrated into NaRPA's current C++ framework to allow for improved realism in light transport simulations.
Second, The current version lacks GPU acceleration, limiting its ability to handle large meshes and render detailed imagery efficiently. Future versions aim to utilize NVIDIA's CUDA and OptiX frameworks to overcome this. 
Additionally, to create simulations that closely mimic spacecraft observations, NaRPA plans to integrate NASA's SPICE\footref{fn:SPICE} toolkit. This integration will enable support for closed-loop simulations using real-time spacecraft positions and lighting angles, significantly enhancing mission-accurate simulations in a navigation pipeline.

\section{Conclusions} \label{sec:s6_conclusion}


In aerospace applications, the testing and validation of vision-based navigation are governed by the ability to generate realistic simulations. 
The Navigation and Rendering Pipeline for Astronautics (NaRPA) is presented in this paper for highlighting visual data generation using physically-valid illumination models for spacecraft navigation.
Capabilities to simulate both space-to-space and ground-to-space observations that are sensitive to atmospheric scattering phenomena are incorporated into the renderer. 
The utility of a multi-sensor visual simulation platform for future spacecraft missions is demonstrated through the ability to generate images, point clouds, depth maps, and contour maps.
Emphasis is placed on validation studies for a velocimeter LiDAR sensor developed for optical simulation studies in entry, descent, and landing applications.
Statistical similarities between the synthetic and realistic sensor data are shown through open-loop validation studies. 
The limitations of NaRPA that must be addressed for high-fidelity sensor simulations for testing and development of spacecraft navigation algorithms are discussed. 
The framework presented in this paper can be easily extended to incorporate custom sensor and shading models that are relevant to any planetary robotic mission. 
Future work will focus on integrating NaRPA with ancillary tools for accurate space-compliant visual navigation simulations.
Sensor models will be thoroughly validated with laboratory experiments for closed-loop simulations of guidance, navigation, and control (GNC) systems for landing and proximity operations around planetary bodies.

\comment{
This research is aimed at rendering physically-accurate computer graphics simulations for space-borne imaging. A rendering engine (NaRPA) is developed to generate illumination models for virtual space-to-space and ground-to-space observations. 

The rendering engine presented in this manuscript incorporates a pipeline for synthesizing images, 3D point clouds, and depth maps from the computer-aided models that describe the virtual scene. The rendering engine determines the intersection of direct and indirect rays with the primitive geometries in the scene. A four-dimensional bidirectional reflectance distribution function is implemented to model the light-matter interaction and determine the shading, i.e., color of the 3D objects.   

The rendering equation is solved by Monte Carlo-based statistical techniques, and photo-realistic images are synthesized by posing the rendering equation as a global illumination problem. To accelerate the evaluation of the rendering equation, the objects in the scene are placed in bounding boxes and arranged in a hierarchical tree structure to enable the computation of fewer ray-surface intersections.

The rendering engine is capable of texture mapping tasks, camera sensor modeling, point cloud, depth, and contour map generation. Furthermore, the engine is capable of creating virtual atmospheric effects which are conducive to ground-to-space observations. An empirically valid refractive index model is adopted to simulate the atmospheric scattering phenomenon. The rendering framework also provides plugins to generate effects of time of day, turbidity, and exposure to replicate real-time observations. 

{The development of NaRPA aims to provide a reliable testing platform for vision-based navigation. It's application to relative navigation problems is emphasized via two applications. First, NaRPA is applied as a stereo image generator for depth estimation using stereoscopy. Using results from a baseline triangulation scheme and point cloud data as ground truth, NaRPA is manifested as a utility for verification of vision algorithms.}

Second, a differentiable rendering technique for vision-based navigation is presented. The technique utilizes the capability of the rendering engine to generate images on the fly given a description of the 3D scene. From the images and the 3D spatial data of an object, six degree of freedom relative pose of the vehicle is iteratively estimated by minimizing the reprojection error in a perspective-n-point model. A proximity operation involving a spacecraft docking sequence with the International Space Station (ISS) is presented to validate the algorithm.
}

\comment{
the various capabilities of the NaRPA framework to imitate realistic observations. It is a ray-tracing based global illumination algorithm to account for a comprehensive nature of light propagation for photo-realism. The framework offers plugins to provide effects of time of day, turbidity, and exposure to include atmospheric effects in the rendering process. To demonstrate the utility of NaRPA as a reliable simulation environment for vision-based navigation, two applications have been demonstrated. 

In the first application, NaRPA is used as a stereo image generator for stereoscopic depth estimation. It is implemented as a tool for the verification of stereo vision algorithms, using the findings from a fundamental triangulation technique and point cloud data as ground truth. In the second application, a differentiable rendering algorithm is proposed for reasoning about 3D scenes from their corresponding 2D projections. In this application,  relative pose is estimated from 3D scenes via nonlinear optimization by propagating gradients of the rendering process through the NaRPA simulations. Furthermore, using representative images and models, NaRPA is demonstrated to be of utility in testing and developing future navigation and guidance missions.}

    \comment{
    The aim of this work is to introduce a physically-based rendering engine capable of producing physics inspired computer graphics simulations \comment{for electro-optical sensors} for aerospace applications. To create illumination models for simulated space-to-space and ground-to-space observations, the Navigation and Rendering Pipeline for Astronautics (NaRPA) is developed. NaRPA uses a ray-tracing pipeline to synthesize pictures, 3D point clouds, and depth maps from computer-aided models that define the virtual scene.
    
    By posing the rendering equation as a global illumination problem, Monte Carlo-based statistical methods are used to solve the rendering equation. The interaction between light and matter is modeled using a four-dimensional bidirectional reflectance distribution function, which also establishes the shading, or color, of the three-dimensional objects. The scene's objects are contained within bounding volumes and grouped in a hierarchical tree structure to enable the computation of fewer ray-surface crossings, which speeds up function evaluations of the rendering equation. The primitive geometries in the scene are what the rendering engine uses to identify where direct and indirect rays intersect.
    
    NaRPA has the capability to model camera sensors in order to generate images, point clouds, depth, and contour maps. Additionally, the engine has the ability to provide simulated atmospheric phenomena that are advantageous for ground-to-space observations. To replicate the atmospheric scattering phenomena, a refractive index model is used that is empirically proven to work. To mimic real-time observations, the rendering framework also offers plugins to provide effects of time of day, turbidity, and exposure. 
    
    The creation of NaRPA aims to offer a trustworthy testing environment for vision-based navigation. Two applications are highlighted to showcase the utility of NaRPA for general navigation. In the first application, NaRPA is used as a stereo image generator for stereoscopic depth estimation. It is implemented as a tool for the verification of stereo vision algorithms, using the findings from a fundamental triangulation technique and point cloud data as ground truth. The second application is a differentiable rendering method for vision-based navigation. The method makes use of the rendering engine's ability to create graphics instantly from a description of a 3D scene. Differentiable rendering propagates the gradients of the image error through simulations. The six degree of freedom relative pose of the vehicle is repeatedly calculated from the images and the 3D spatial data of features by minimizing the reprojection error in a perspective-n-point model. To test the capability of NaRPA, a spacecraft docking procedure with the International Space Station (ISS) is demonstrated in a proximity operation scenario.
    
    In summary, NaRPA is designed to provide a high degree of realism and offers a powerful collection of tools for digitally simulating ground-based and space-based optical sensors and employing them in navigation pipelines. The data obtained are shown to be used for off-line and closed-loop simulations of proximity operations, and terrain-relative navigation. The paper presents several examples concerning optical design simulation in different domains to demonstrate the utility of NaRPA for astronautic applications.
    }

\appendix

\section{Mathematical preliminaries for light matter interactions} \label{sec:rayObjIntersection}

In the ray tracer, the path of light is defined as a ray emanating from the origin of the sensor coordinate system, ${u}-{v}-{w}$. The rendering engine computes intersections between light rays and the 3D objects in the environment. Given a sensor orientation, the first objective is to find the intersection of primary rays (the first rays cast into the environment) with any geometry in the scene. The type of objects in geometric models generally include spheres, triangles, rectangles, and in general polygons. The mathematical expressions for evaluating ray-triangle and ray-sphere intersections are delineated in the literature and adopted for this ray tracer engine~\citep{10.5555/1628957}. Owing to their co-planar nature and efficient software realization, triangular meshes are  adopted as primitive representations that build 3D geometries for models involving space based observations. The ray-primitive intersections are highlighted in Fig.~\ref{fig:rayPrimitiveIntersection} and their mathematical preliminaries are derived as follows.

\begin{figure}[ht]
  \centering
  \subfloat[The ray $\mathbf{p}(t)$ hits the sphere at points $p_1$ and $p_2$]{\includegraphics[width=0.28\textwidth]{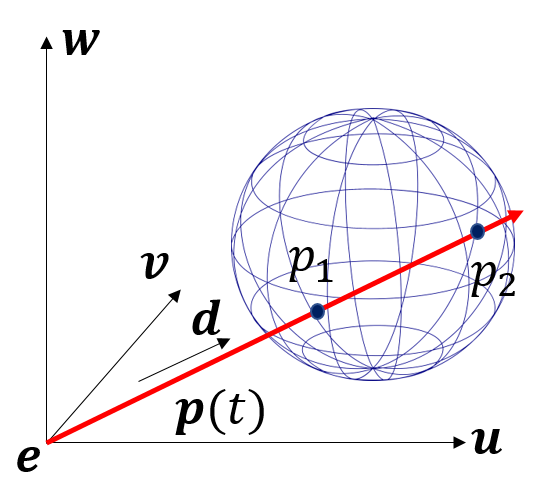}\label{fig:raySphere}}
  \hfill
  \subfloat[The ray hits the plane containing the triangle at point $\mathbf{p}$] {\includegraphics[width=0.21\textwidth]{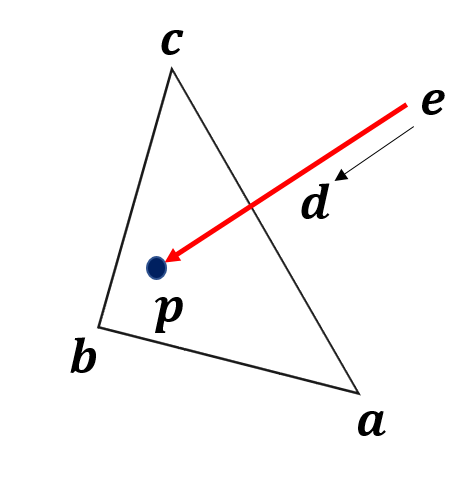}\label{fig:rayTriangle}}
  \caption{Intersection of a ray with a sphere and a triangle.}
  \label{fig:rayPrimitiveIntersection}
\end{figure}

A ray $\mathbf{p}(t)$ can be mathematically described using the parametric equation of a line originating from the origin of the sensor coordinate system, $\mathbf{e}$ and advancing a distance of $\mathit{t}$ units in the direction $\mathbf{d}$ 
\begin{equation} \label{eq:1}
    \mathbf{p}(t) = \mathbf{e} + \mathit{t}\mathbf{d}
\end{equation}
\noindent The point of intersection of this ray, $\textbf{p}(t)$, with a sphere centered at $\textbf{c}$ and having radius $R$ is obtained by solving the following equation:
%
\begin{equation} \label{eq:2}
    (\mathbf{p} - \mathbf{c}) . (\mathbf{p} - \mathbf{c}) - R^2 = 0
\end{equation}
\noindent Since the point at the intersection lies on both rays in Eq. (\ref{eq:1}) and the sphere in Eq. (\ref{eq:2}), the values of $\mathit{t}$ on the ray are solved by substitution. This yields:
\begin{equation}
(\mathbf{e} + \mathit{t}\mathbf{d} - \mathbf{c}) . (\mathbf{e} + \mathit{t}\mathbf{d} - \mathbf{c}) - R^2 = 0
\end{equation}
\begin{equation} \label{eq:quadraticRay}
    (\mathbf{d}.\mathbf{d})\mathit{t}^2 + 2\mathbf{d}.(\mathbf{e} - \mathbf{c})\mathit{t} +  (\mathbf{e} - \mathbf{c}) . (\mathbf{e} - \mathbf{c}) - R^2 = 0
\end{equation}
\noindent  The roots of the quadratic equation in Eq. (\ref{eq:quadraticRay}) determine the point of ray-sphere intersection. If the discriminant of the Eq. (\ref{eq:quadraticRay}) is negative, then there are no intersections between the sphere and the ray. Conversely, a positive discriminant implies that there exist two solutions, each of which describes the points on the sphere where the ray enters and where the ray exits, respectively. If the discriminant is zero, then the ray grazes the surface of the sphere and touches it at exactly one point. Solving for the parameter $\mathit{t}$, we get the intersection point of the ray sphere defined by 
\begin{equation}
    \mathit{t} = \frac{-\mathbf{d}. (\mathbf{e} - \mathbf{c}) \pm \sqrt{(\mathbf{d}.( \mathbf{e} - \mathbf{c}) - (\mathbf{d}.\mathbf{d}) ((\mathbf{e} - \mathbf{c}) . (\mathbf{e} - \mathbf{c}) - R^2 )}}{(\mathbf{d}.\mathbf{d})}
\end{equation}

\noindent The unit surface normal vector $\mathbf{n}$ at the ray-sphere intersection point is given by $\mathbf{n} = {(\mathbf{p}-\mathbf{c})}/{R}$. The angle between the light direction and the surface normal vector determines the spectral reflection at the intersection point, toward the sensor.

The intersection of a ray with a triangle may be computed using M{\"o}ller-Trumbore algorithm, which uses barycentric coordinates and takes advantage of the parametric equation for a plane containing the triangle~\citep{moller1997fast}. The barycentric coordinates represent the intersection point $\mathbf{p}$ in terms of the non-orthogonal basis vectors formed by the vertices $\mathbf{a}, \mathbf{b}$ and $\mathbf{c}$ of a triangle as follows:
\begin{equation} \label{eq:6}
    \mathbf{p}(\beta,\gamma) = \mathbf{a} + \beta (\mathbf{b} - \mathbf{a}) + \gamma (\mathbf{c} - \mathbf{a})    
\end{equation}
\noindent The intersection point, $\mathbf{p}$, is obtained by replacing the $\mathbf{p}$ with the parametric form of the ray in Eq. (\ref{eq:1}) as follows: 
\begin{equation} \label{eq:7}
    \mathbf{e}+t\mathbf{d} = \mathbf{a} + \beta (\mathbf{b} - \mathbf{a}) + \gamma (\mathbf{c} - \mathbf{a})
\end{equation}
\noindent In this representation using the barycentric coordinate system\comment{\cite{weisstein2003barycentric}}, the point $\mathbf{p}$ is within the triangle only if $\beta > 0$, $\gamma > 0$ and ${\beta} + {\gamma} < 1$. Equation \ref{eq:7} is transformed into a linear system of equations and solved for $t$, $\beta$, and $\gamma$ using Cramer's rule or matrix inversion.  
\begin{equation} \label{eq:8}
    \begin{bmatrix} \mathbf{a-b} & \mathbf{a-c} & \mathbf{d} \end{bmatrix} \begin{bmatrix}
    \beta \\ \gamma \\ t
    \end{bmatrix} = \begin{bmatrix} \mathbf{a-e} \end{bmatrix}
\end{equation}

\section*{Declaration of Competing Interest}
The authors declare that they have no known competing financial interests or personal relationships that could have appeared to influence the work reported in this paper.

\section*{Acknowledgments}
This research was carried out at Texas A\&M University and the Jet Propulsion Laboratory, California Institute of Technology, under a contract with the National Aeronautics and Space Administration (80NM0018D0004). 
Support from JPL’s Strategic University Research Partnership (SURP) initiative is gratefully acknowledged.  
Drs. Carolina Restrepo (NASA-GSFC), Jerry Tessendorf (Clemson University), Daniel Crispell (Vision Systems Inc.), John Crassidis (University at Buffalo), Sarah Stevens (JPL), Anup Katake (JPL), Alejandro M San Martin (JPL), Nicolas Trawney (JPL), Eli D. Skulsky (JPL), Tejas Kulkarni (JPL), Georgios Georgakis (JPL), Daniel Crispell (VSI), Joe Mundy (VSI), Mr. Ronney Lovelace (NASA-JSC), and Mr. David van Wijk (TAMU), Dr. Caleb Peck (TAMU), Dr. James McElreath (TAMU) are acknowledged for their inputs and support in this development. 

\bibliographystyle{jasr-model5-names}
\biboptions{authoryear}
\bibliography{sections/References,sections/refsNew}

\end{document}